\documentclass[twocolumn,pra,twocolumn,superscriptaddress]{revtex4}
\usepackage{color}
\usepackage{amsmath}
\usepackage{amssymb}
\usepackage{graphicx}
\usepackage{tikz}

\usepackage{subfigure}

\usepackage{float}
\usepackage{bbm}

\usepackage{epsfig} 
\usepackage{verbatim}
\usepackage{array}
\usepackage{setspace}

\usepackage[unicode=true,
 bookmarks=true,bookmarksopen=false,
 breaklinks=false,pdfborder={0 0 1},backref=false,colorlinks=true]
 {hyperref}
\hypersetup{
 linkcolor=magenta, urlcolor=blue, citecolor=blue, pdfstartview={FitH}, hyperfootnotes=false, unicode=true}

\makeatletter
\@ifundefined{textcolor}{}
{%
 \definecolor{BLACK}{gray}{0}
 \definecolor{WHITE}{gray}{1}
 \definecolor{RED}{rgb}{1,0,0}
 \definecolor{GREEN}{rgb}{0,1,0}
 \definecolor{BLUE}{rgb}{0,0,1}
 \definecolor{CYAN}{cmyk}{1,0,0,0}
 \definecolor{MAGENTA}{cmyk}{0,1,0,0}
 \definecolor{YELLOW}{cmyk}{0,0,1,0}
}

\usepackage{amsfonts}\usepackage{tabularx}\usepackage{dcolumn}\usepackage{bm}\usepackage{graphicx}\usepackage{epstopdf}

\hypersetup{urlcolor=blue}

\makeatother

\newcolumntype{C}[1]{>{\centering\arraybackslash$}p{#1}<{$}}

\begin{document}

\title{Exploring Entanglement Spectrum and Phase Diagram in multi-electron Quantum Dot Chains
}

\author{Guanjie He}
\affiliation{Department of Physics, City University of Hong Kong, Tat Chee Avenue, Kowloon, Hong Kong SAR, China, and City University of Hong Kong Shenzhen Research Institute, Shenzhen, Guangdong 518057, China}
\author{Xin Wang}
\email{x.wang@cityu.edu.hk}
\affiliation{Department of Physics, City University of Hong Kong, Tat Chee Avenue, Kowloon, Hong Kong SAR, China, and City University of Hong Kong Shenzhen Research Institute, Shenzhen, Guangdong 518057, China}

\date{\today}

\begin{abstract}

We investigate the entanglement properties in semiconductor quantum dot systems modeled by extended Hubbard model, focusing on the impact of potential energy variations and electron interactions within a four-site quantum dot spin chain. Our study explores local and pairwise entanglement across configurations with electron counts $N=4$ and $N=6$, under different potential energy settings. By adjusting the potential energy in specific dots and examining the entanglement across various interaction regimes, we identify significant variations in the ground states of quantum dots. Our results reveal that local potential modifications lead to notable redistributions of electron configurations, significantly affecting the entanglement properties. These changes are depicted in phase diagrams that show entanglement dependencies on interaction strengths and potential energy adjustments, highlighting complex entanglement dynamics and phase transitions triggered by inter-dot interactions.


\end{abstract}

\maketitle

\section{introduction}

Quantum entanglement plays a crucial role in various fields of quantum physics, including quantum communication and quantum information processing \cite{RevModPhys.77.513, Abaach2023}. In condensed matter physics, especially in many-body quantum systems, quantum entanglement serves as a fundamental criterion for quantum phase transitions and many-body localization \cite{RevModPhys.80.517,PhysRevLett.122.040606,PhysRevB.87.134202,PhysRevB.82.174411}. Among various systems, semiconductor quantum dots have emerged as scalable, implementable, and precisely controllable \cite{PRXQuantum.2.040306, Gonzalez-Zalba2021, Philips2022, RevModPhys.95.011006,PhysRevLett.116.110402, Feng2021, PhysRevB.107.085427, PhysRevLett.108.140503} platforms for simulating  many-body systems of interest, in particular the Fermi-Hubbard physics \cite{10.1038/nature23022, PhysRevX.11.041025, PhysRevB.107.014403, Dehollain2020, Kiczynski2022, Wang2022, Le2020}. The Fermi-Hubbard model provides a common framework for describing quantum dot systems in the regime of low temperatures and strong Coulomb interactions, finding extensive applications in the physical realization of quantum information processing \cite{PhysRevB.84.115301,PhysRevB.83.235314,PhysRevB.83.161301,Watson2018}. Consequently, a comprehensive understanding of quantum dots from the perspective of Fermi-Hubbard physics becomes imperative.

High-fidelity qubit gate operations \cite{doi:10.1126/science.aao5965, 10.1038/s41467-020-17865-3} and noise suppression schemes \cite{PhysRevLett.116.116801} commonly applied to conventional quantum dot systems, where each dot accommodates at most two electrons, traditionally rely on the monotonically increasing behavior of exchange energy as a function of detuning \cite{PhysRevA.57.120, PhysRevB.59.2070, PhysRevB.82.045311, maune2012coherent, PhysRevB.90.195424}. However, recent investigations \cite{PhysRevLett.119.227701, PhysRevX.8.011045, PhysRevB.84.235309, PhysRevLett.114.226803, Leon.21, PhysRevB.104.L081409, PhysRevLett.127.086802, PhysRevLett.112.026801, PhysRevB.97.245301, PhysRevB.105.245409, PhysRevA.64.042312,ercan2023MULTI} have revealed the interesting properties of specific quantum dots capable of hosting more than two electrons, such as non-monotonic behavior of exchange energy with distinct sweet spots \cite{PhysRevB.105.245409,PhysRevB.106.075417}, fast spin exchange dynamics \cite{10.1038/s41467-019-09194-x}, superexchange interactions between non-neighboring dots \cite{PhysRevB.102.035427, PhysRevA.106.022420, PhysRevLett.126.017701}, and resilience to noise \cite{PhysRevB.91.155425, PhysRevB.88.161408, PhysRevB.106.075417, qute.202200074}. These properties can be attributed to the influence of higher excited orbitals and can be effectively understood within the framework of the Full Configuration Interaction \cite{10.1063/1.2179418} and the extended Hubbard Model (EHM), which incorporates multiple energy levels. 

The entanglement spectrum of the one-dimensional EHM in its ground state has been well-understood \cite{PhysRevB.105.115145, PhysRevB.92.075423, PhysRevB.75.165106, PhysRevLett.93.086402}. Consequently, in the case of a half-filled system, the entanglement properties of a quantum dot spin chain can be effectively explained \cite{PhysRevA.106.022421}. However, when there is a tilted potential energy difference among the dots, the mirror symmetry of the system is broken, which leads to the tunable entanglement values through the application of precise electron control using external electric fields \cite{PhysRevB.101.045306}. These previous works have motivated us to investigate the entanglement spectrum of a quantum dot spin chain where each dot incorporates multiple energy levels. This exploration holds great potential for uncovering the rich physical properties of quantum dot systems.
 
In this study, we investigate the entanglement patterns of the ground states of multi-electron quantum dot systems using the EHM, which incorporates multiple orbitals within each dot. Our specific focus lies in characterizing the entanglement properties of one-site and two-site reduced density matrices. By computing and analyzing the entanglement spectrum for various system sizes, we uncover notable findings. Firstly, when there are no potential energy differences among the dots, the multi-electron quantum dot system can be accurately described by the EHM, either in a half-filled state or a non-half-filled state, depending on the total electron number. However, when a selected dot within the chain exhibits a potential energy difference relative to its neighboring dots, distinct system phases and phase boundaries emerge in the entanglement spectrum. These phases depend on the coupling strengths and potential energy difference values. The emergence of these phases indicates that the presence of a selected dot with a potential energy ladder profoundly impacts the electron configuration in its vicinity. This influence is more pronounced in small systems while limited in larger-size systems, due to the size effect.

This paper is organized as follows. In Section~\ref{sec:model}, we present the EHM as a suitable framework for describing multi-electron quantum dot chain systems. Section~\ref{sec:entanglement} introduces the definition of one-site and two-site reduced density matrices and entanglement entropy for these systems. Our main results are presented in Section~\ref{result}, starting with an examination of a system size of $L=4$ and electron numbers $N=4$ and $N=6$. We analyze the entanglement spectrum properties with and without potential energy differences. Furthermore, we extend our analysis to larger system sizes as $L$ approaches infinity. Finally, we summarize our findings and provide concluding remarks in Section~\ref{sec:conclusion}.

\section{Extended Hubbard Model}

\begin{figure}
	(a)\includegraphics[scale=0.16]{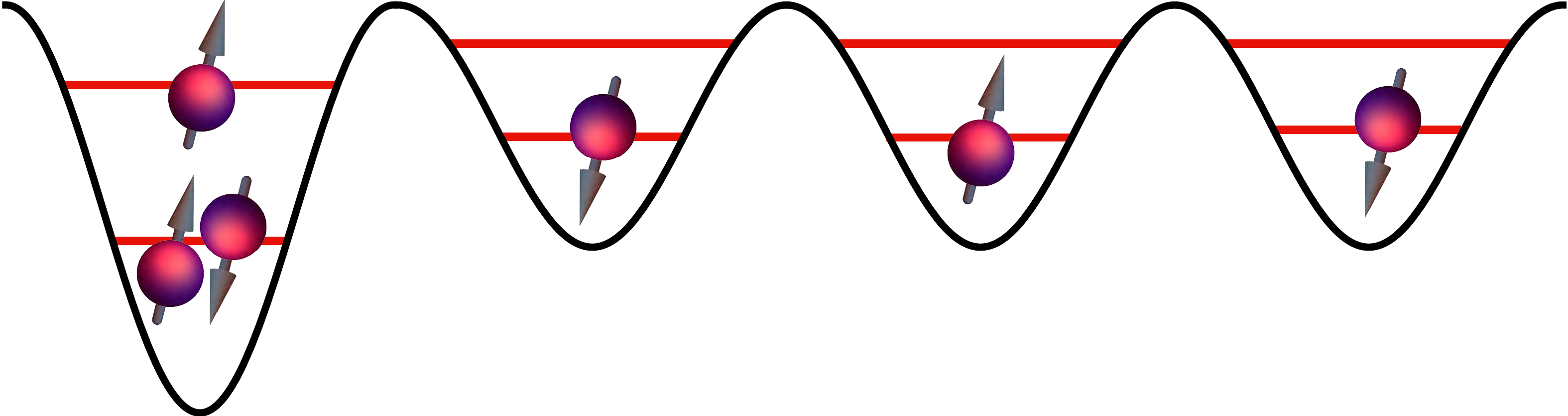}
	(b)\includegraphics[scale=0.32]{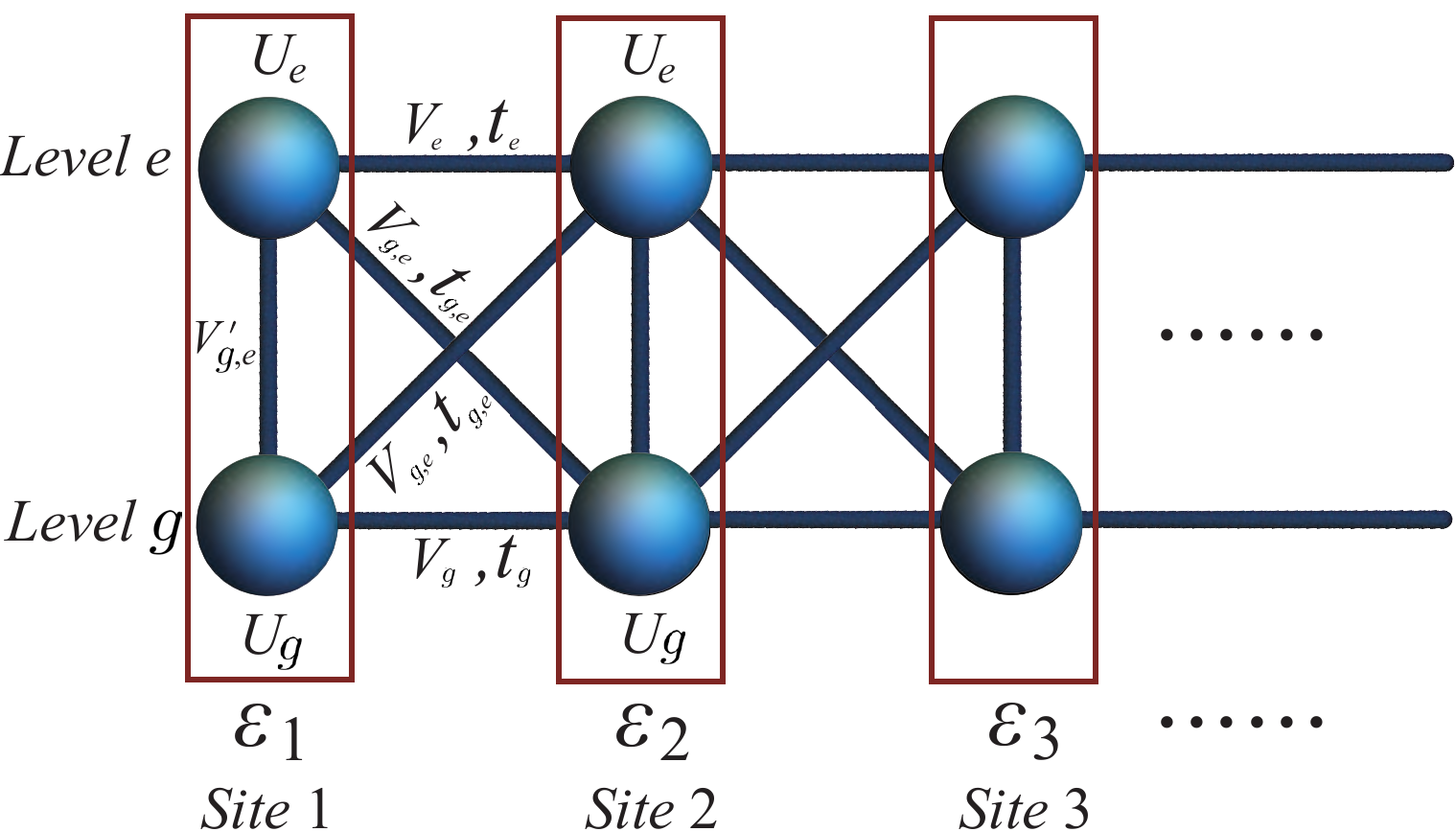}
	\caption{(a) Schematic illustration of a $L=4$ multi-electron quantum dot spin chain system hosting $N=6$ electrons. (b) In two-level case, the equivalent asymmetric Hubbard ladder described by Hamiltonian~\eqref{eq:Hubbard}. The box indicates for each site $i$, electrons on different energy levels have the same detuning energy $\varepsilon _{i}$. $g$ indicates the ground level and $e$ the first excited level.}\label{fig:Fig1dotmuti}
\end{figure}
\label{sec:model}

We consider a Multiple-Quantum-Dot system (MQD) (schematically shown in Fig.~\ref{fig:Fig1dotmuti}), described by an EHM with short-range Coulomb interactions and tunneling restricted to nearest-neighbor sites within the same energy level and the nearest-neighbor energy level. The model can be described by the following Hamiltonian:

\begin{equation}
	\label{eq:Hubbard}
	\begin{split}
		H=&-\sum_{i,\nu,\overline{\nu},\sigma}(t_{\nu}c_{i,\nu,\sigma}^{\dagger}c_{i+1,\nu,\sigma}+t_{\nu,\overline{\nu}}c_{i,\nu,\sigma}^{\dagger}c_{i+1,\overline{\nu},\sigma}+\mathrm{H.c.})\\
		&+\sum_{i,\nu,\overline{\nu},\sigma}(V_{\nu}n_{i,\nu,\sigma}n_{i+1,\nu,\sigma'}+V_{\nu,\overline{\nu}}n_{i,\nu,\sigma}n_{i+1,\overline{\nu},\sigma'}\\
		&+V'_{\nu,\overline{\nu}}n_{i,\nu,\sigma}n_{i,\overline{\nu},\sigma'})+\sum_{i,\nu}U_{\nu}n_{i,\nu\downarrow}n_{i,\nu\uparrow}\\
		&+\sum_{i,\sigma}\varepsilon _{i,\sigma}n_{i\sigma},
	\end{split}
\end{equation}
where $i$ indicates the quantum dot site, $\nu$ and $\overline{\nu}$ denotes different orbital level, which can be either ground orbital ($g$) or excited orbital ($e$), while $\sigma$ and $\sigma'$ refer to the spins that can be either up or down. $\varepsilon _{i,\sigma}$ is the potential energy at dot $i$, note that although in one quantum dot, electrons can occupy different orbitals, they share the same potential energy. $t_{\nu}$ is the tunneling energy between $i$th and $(i+1)$th site at $\nu$th orbital level, $t_{\nu,\overline{\nu}}$ is the tunneling energy between $i$th site at $\nu$th orbital level and  $(i+1)$th site at $\overline{\nu}$th orbital level, i.e. $t_{g,e}$ or $t_{e,g}$. $U_{\nu}$ denotes the on-site Coulomb interaction in the $\nu$th orbital level, $V_{\nu}$ is the nearest direct Coulomb interaction between the $i$th and $(i+1)$th site at $\nu$th orbital, $V_{\nu,\overline{\nu}}$ is the nearest direct Coulomb interaction between the $i$th site at $\nu$th orbital and $(i+1)$th site at $\overline{\nu}$th orbital, and finally, $V'_{\nu,\overline{\nu}}$ is the nearest direct Coulomb interaction between the $\nu$th orbital and $\overline{\nu}$th orbital at $i$th site, i.e. $V_{g,e}$, $V_{e,g}$, $V'_{g,e}$, $V'_{e,g}$.

According to the Pauli exclusion principle, electrons have four occupation states $|v\rangle_{i,\nu}$=$|0 \rangle_{i,\nu}$, $|\uparrow \rangle_{i,\nu}$, $|\downarrow \rangle_{i,\nu}$, $|\uparrow\downarrow \rangle_{i,\nu}$ in the $\nu$th orbital of the $i$th site. Thus, the dimension of the Hilbert space for an $L$-site MQD chain with $K$ orbitals per site is $4^{LK}$. The configuration basis states are $|v_1,v_2,...,v_L\rangle$ = $\prod_{i=1}^{L}|v_i\rangle_i$, where $|v_i\rangle_i$ = $\prod_{\nu=1}^{K}|v\rangle_{i,\nu}$ represents the configuration basis for the i-th site. In this work, we numerically study MQD chains with $N$ and $N+2$ electrons in an $L=N$ sites systems, restricting our analysis to the ground and first excited orbital states ($\nu=g,e$) for each quantum dot.

\section{Reduced density matrices and Entanglement}

\begin{figure*}
	(a)\includegraphics[scale=0.33]{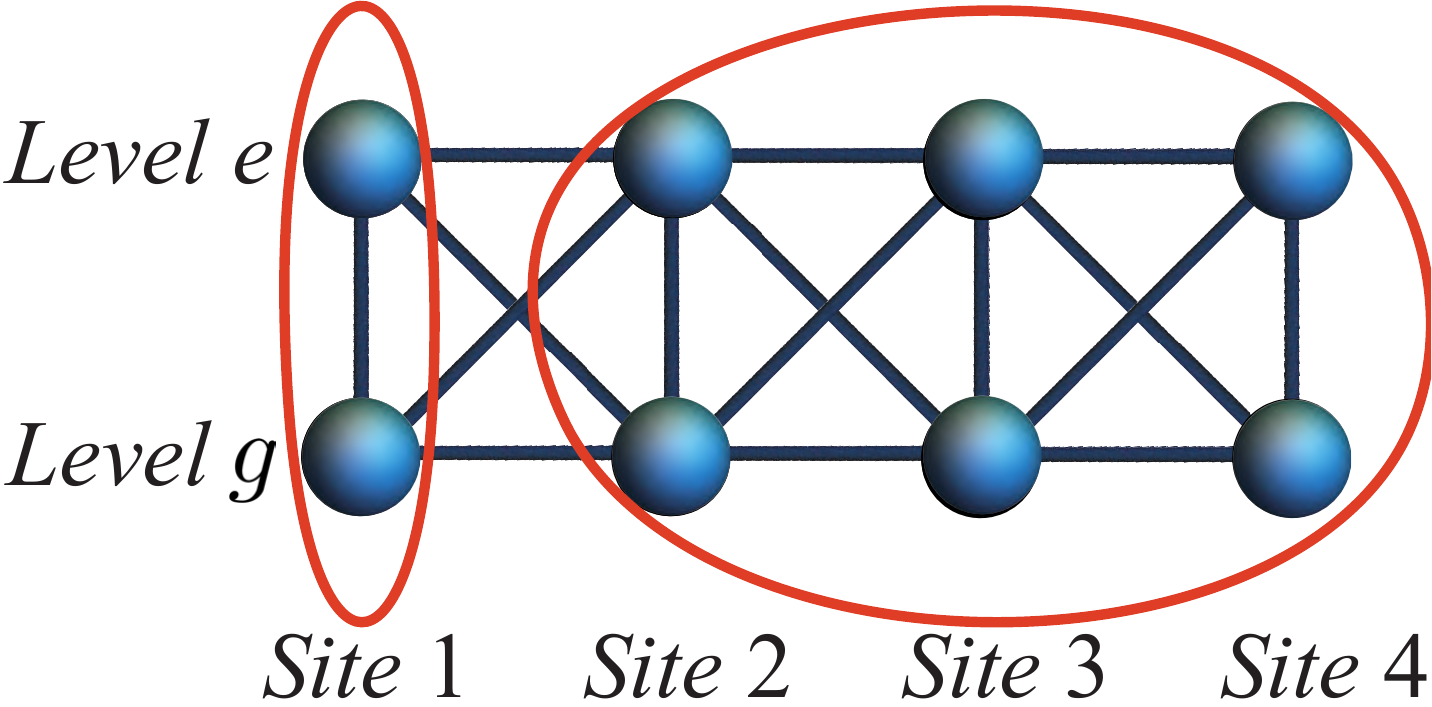}
	(b)\includegraphics[scale=0.33]{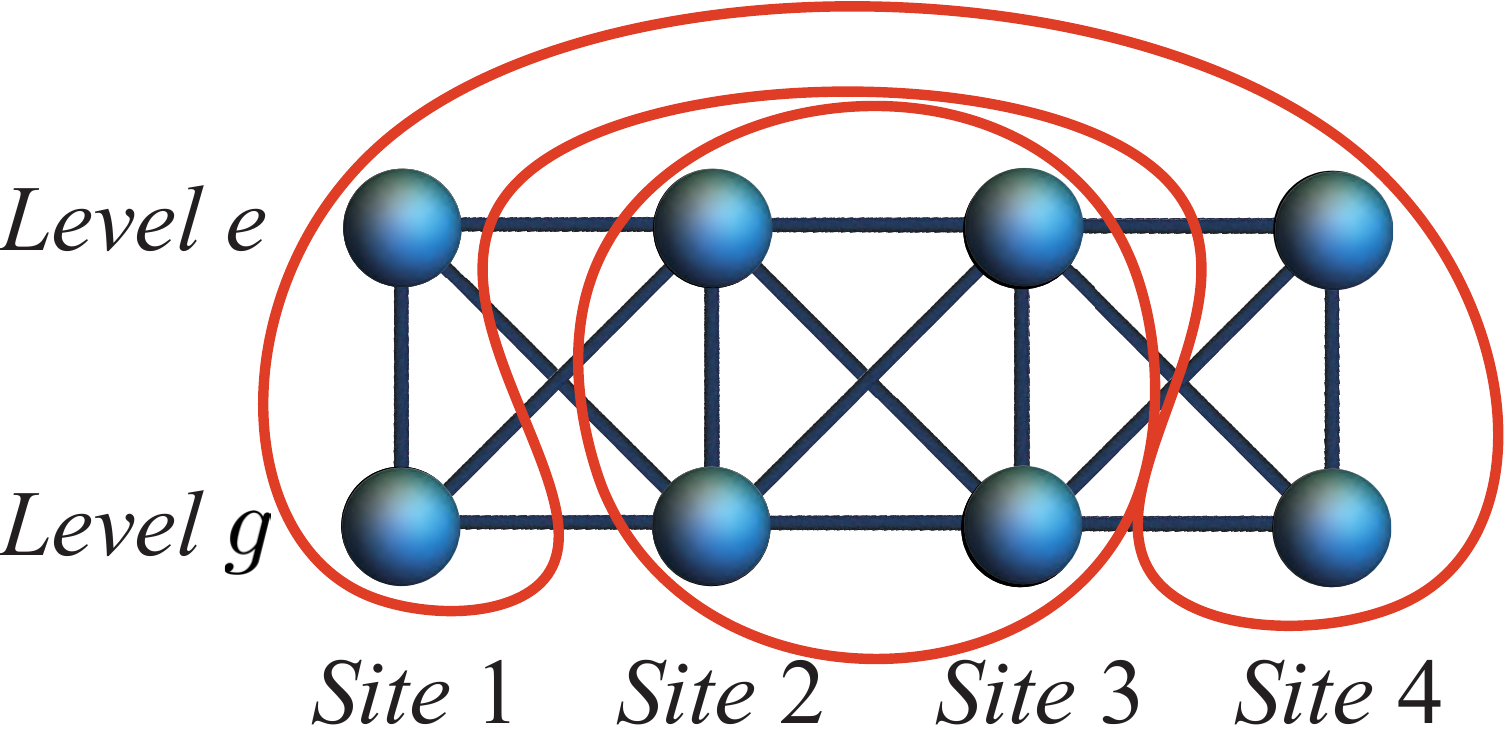}
	\caption{Illustration of bipartite entanglement and quantum states in $L=4$ two-level system (a) Local entanglement $E(\rho_{1})$ and (b) Pairwise entanglement $E(\rho_{14})=E(\rho_{23})$. The red circle indicates the selected partition. $g$ indicates the ground level and $e$ the first excited level.}\label{fig:entanglement}
\end{figure*}

\label{sec:entanglement}

We first obtain the ground state (GS) $|\psi_{\mathrm{GS}}\rangle$ of the system by diagonalizing the Hamiltonian. The GS can be expressed as a linear superposition of all possible electron configuration basis states $|\psi_{m}\rangle$ in the occupation number representation $|v_{1},v_{2}...v_{L}\rangle$:
\begin{equation}
	|\psi_\mathrm{GS}\rangle=\sum_{m}c_{m}|\psi_{m}\rangle,
\end{equation}
where $c_m$ are the coefficients of the superposition.

The density matrix $\rho_\mathrm{GS}$ of the entire system can be expressed as a sum of the occupation probabilities $P_m$ of all electron configurations $|\psi_{m}\rangle$:
\begin{equation}
	\rho_\mathrm{GS} =\sum_{m}P_{m}|\psi_{m} \rangle\langle\psi_{m}|.
\end{equation}

To analyze entanglement, we divide the full system into subsystems A and B. The reduced density matrix $\rho_A$ for subsystem A is obtained by taking the partial trace of $\rho_\mathrm{GS}$:
\begin{equation}
	\rho_{A} =\textrm{Tr}_{B}\rho_\mathrm{GS}.
\end{equation}

The von Neumann entropy $E(\rho_{A})$ measures the entanglement between subsystem A and the remaining subsystem B, and is defined as:
\begin{equation}
	E(\rho_{A}) =-\textrm{Tr}(\rho_{A} \log_2 \rho_{A}).
\end{equation}

\subsection{Local Entanglement of multi-electron quantum dot}\label{subsec:localent}

In this paper, we focus on GaAs QD, since silicon QD is more complicated due to the valley splitting and phases \cite{Tariq2022}. In GaAs QD \cite{PhysRevB.84.235309},  in the two orbitals and within the parameters we considered, electrons prefer to doubly occupy ground states before filling the first excited states. Therefore the state space of a single site is spanned by nine bases: $\{ |0,0\rangle, |\uparrow_{g}$,$0\rangle,|\downarrow_{g},0\rangle, |\uparrow_{g}\downarrow_{g},0\rangle,|\uparrow_{g},\downarrow_{e}\rangle,|\downarrow_{g},\uparrow_{e}\rangle, |\uparrow_{g}\downarrow_{g},\uparrow_{e}\rangle, |\uparrow_{g}\downarrow_{g},\downarrow_{e}\rangle, |\uparrow_{g}\downarrow_{g}$, $\uparrow_{e}\downarrow_{e}\rangle  \} $. $0_{g}$ and $0_{e}$ represent cases with no electron occupying the ground and the first excited orbital, respectively. $\uparrow_{g}$, $\downarrow_{g}$, $\uparrow_{e}$, $\downarrow_{e}$ stand for an electron with spin up or down indicated as the arrow staying in the ground $(g)$ and the first excited orbital $(e)$ indicated in the subscript respectively.

The two-level one-site reduced density matrix for site $i$ can be written as
\begin{equation}
\rho_{i}=\textrm{Tr}_{i}(\rho_\mathrm{GS}).
\end{equation}
Expressing in terms of basis: $\{ |0,0\rangle, |\uparrow_{g}$,$0\rangle,|\downarrow_{g},0\rangle, |\uparrow_{g}\downarrow_{g},0\rangle,|\uparrow_{g},\downarrow_{e}\rangle,|\downarrow_{g},\uparrow_{e}\rangle, |\uparrow_{g}\downarrow_{g},\uparrow_{e}\rangle, |\uparrow_{g}\downarrow_{g},\downarrow_{e}\rangle, |\uparrow_{g}\downarrow_{g}$, $\uparrow_{e}\downarrow_{e}\rangle\} $, $\rho_{i}$ can be written as a $9\times 9$ matrix as follows:

 $\rho_{i}$=$\begin{pmatrix}
	v_{i,1}&  &  &  &  &  &  &  & \\ 
	&  v_{i,2}&  &  &  &  &  &  & \\ 
	&  &  v_{i,3}&  &  &  &  &  & \\ 
	&  &  &  v_{i,4}&  v_{i,a}&  v_{i,b}&  &  & \\ 
	&  &  &  v_{i,a}&  v_{i,5}&  v_{i,c}&  &  & \\ 
	&  &  &  v_{i,b}&  v_{i,c}&  v_{i,6}&  &  & \\ 
	&  &  &  &  &  &  v_{i,7}&  & \\ 
	&  &  &  &  &  &  &  v_{i,8}& \\ 
	&  &  &  &  &  &  &  &v_{i,9} 
\end{pmatrix}$.
Here, $v_{i,m}$($m=1,2,...,9$), $v_{i,a}$, $v_{i,b}$ and $v_{i,c}$ are determined by potential energy $\varepsilon$ of different dots and quantity $U$. In half-filled case, when there is no potential energy difference of all quantum dots, the local reduced density matrix $\rho_{i}$ can be simplified to one energy level case \cite{PhysRevLett.93.086402}, with
\begin{subequations}
\begin{align}
	v_{i,1}=1-v_{i,4}+v_{i,2}+v_{i,3},\\
    v_{i,2}=\left \langle n_{i,g,\uparrow} \right \rangle-v_{i,4},\\
	v_{i,3}=\left \langle n_{i,g,\downarrow} \right \rangle-v_{i,4},\\
	v_{i,4}=\textrm{Tr}(n_{i,g,\uparrow}n_{i,g,\downarrow}\rho_{i})=\left \langle n_{g\uparrow}n_{g\downarrow} \right \rangle,\\
	v_{i,a}=v_{i,b}=v_{i,c}=0,\\
	v_{i,5}=v_{i,6}=v_{i,7}=v_{i,8}=v_{i,9}=0.
\end{align}
\end{subequations}

When potential energy differences exist between quantum dots (in particular, in our work, only one site's potential energy is altered while the remaining sites have no potential energy difference), the contributions of $v_{i,5}$, $v_{i,6}$, $v_{i,7}$, $v_{i,8}$, $v_{i,9}$, $v_{i,a}$, $v_{i,b}$ and $v_{i,c}$ cannot be ignored. Therefore, the above expression of $v_{i,m}$ does not hold. However, we can still derive that $v_{i,2}=v_{i,3}$, $v_{i,5}=v_{i,6}$, and $v_{i,7}=v_{i,8}$. In particular, for GaAs, within the parameters we set (which will be explained in detail later), the basis of $|\uparrow_{g},\downarrow_{e}\rangle$ and $|\downarrow_{g},\uparrow_{e}\rangle$ are energetically unfavorable and therefore have no contribution, leading to $v_{i,5}, v_{i,6}, v_{i,a}, v_{i,b}, v_{i,c}\sim0$ at any potential $V_{i}$. Thus, $\rho_{i}$ can be represented as a $7 \times 7$ diagonal matrix as:

\begin{equation}
	\begin{split}
		\rho_{i}=&v_{i,1}|0_{g},0_{e}\rangle \langle0_{g},0_{e}|+v_{i,2}|\uparrow_{g},0_{e}\rangle \langle\uparrow_{g},0_{e}|\\
		&+v_{i,3}|\downarrow_{g},0_{e}\rangle \langle\downarrow_{g},0_{e}|+v_{i,4}|\uparrow_{g}\downarrow_{g},0_{e}\rangle \langle\uparrow_{g}\downarrow_{g},0_{e}|\\
		&+v_{i,7}|\uparrow_{g}\downarrow_{g},\uparrow_{e}\rangle \langle\uparrow_{g}\downarrow_{g},\uparrow_{e}|+v_{i,8}|\uparrow_{g}\downarrow_{g},\downarrow_{e}\rangle \langle\uparrow_{g}\downarrow_{g},\downarrow_{e}|\\
		&+v_{i,9}|\uparrow_{g}\downarrow_{g},\uparrow_{e}\downarrow_{e}\rangle \langle\uparrow_{g}\downarrow_{g},\uparrow_{e}\downarrow_{e}|.
	\end{split}
\end{equation}

For the $N=4$ system, there are four distinct approaches to analyzing local bipartite entanglement: $E(\rho_{1})$, $E(\rho_{2})$, $E(\rho_{3})$ and $E(\rho_{4})$. An example of this can be seen in Fig.~\ref{fig:entanglement}(a), which shows the local entanglement $E(\rho_{1})$.

\subsection{Pairwise Entanglement of multi-electron quantum dot}

Similarly, for site $i$ and site $j$, the two-site reduced density matrix can be written as 
\begin{equation}
	\rho_{ij}=\textrm{Tr}_{ij}(\rho_\mathrm{GS}).
\end{equation}
As depicted in Fig.~\ref{fig:entanglement}(b). According to the nine bases considered for a single site in Sec.~\ref{subsec:localent}, the electrons in two sites with two orbitals have $9^{2}=81$ possible configurations. With respect to these bases, $\rho_{ij}$ can be described as an $81\times 81$ matrix. Similar to one site case, where we dropped two energetically unfavorable bases $|\uparrow_{g},\downarrow_{e}\rangle$ and $|\downarrow_{g},\uparrow_{e}\rangle$, $\rho_{ij}$ can be described as a $49\times 49$ matrix since electrons in two sites have $7^{2}=49$ occupation probabilities. There are three possible approaches to analyzing pairwise bipartite entanglement for the $N=4$ system, : $E(\rho_{12})$ and $E(\rho_{34})$, $E(\rho_{13})$ and $E(\rho_{24})$, $E(\rho_{14})$ and $E(\rho_{23})$. Fig.~\ref{fig:entanglement}(b) demonstrates one possible  bipartite pairwise entanglement $E(\rho_{14})$ and $E(\rho_{23})$.

\section{Results}

\label{result}

In our GaAs quantum dots system setup, we have defined a set of parameters that can represent the properties of multi-electron dots \cite{PhysRevB.97.245301,PhysRevB.105.245409}. Accordingly, we set that the tunneling energy between the nearest sites is larger for lower orbitals, and is smaller for higher orbitals. This means that the tunneling between two ground orbitals is the greatest, followed by the tunneling between one ground orbital and one excited orbital, and finally, the tunneling between two excited orbitals, i.e.,  $t_{e} < t_{g,e} < t_{g}$.  Similarly, within one single dot or between two nearest dots, the on-site Coulomb interaction energy and the nearest direct Coulomb interaction energy from higher orbitals are larger than those from lower energy levels, since the electron that occupies a higher orbital requires more energy, i.e., $U_{g} < V'_{g,e} < U_{e}$ and $V_{g} < V_{g,e} < V_{e}$. The numerical relation between $V_{\nu}$ and $U_{\nu}$ is referenced from \cite{PhysRevB.105.245409, PhysRevB.95.241303, 10.1038/s41534-021-00449-4, PhysRevApplied.12.064049, PhysRevB.83.235314}, satisfying a strong repulsive on-site interaction regime in EHM \cite{ PhysRevLett.93.086402}, i.e., $V_{\nu}<U_{\nu}$ and $V_{g,e}<V'_{g,e}$. In one-dimensional EHM at half filling, the ratio between on-site Coulomb interaction $U_{\nu}$ and the nearest direct Coulomb interaction energy $V_{\nu}$ will lead to charge-density wave (CDW) order and spin-density wave (SDW) in the strong-coupling limit regime \cite{ PhysRevLett.93.086402}.  Specifically, for $U_{g}>2V_{g}$, the ground state is a staggered charge-density-wave, and for $U_{g}<2V_{g}$, the ground state is a staggered spin density wave. These spin order properties will also be apparent in our simulation results due to the chosen parameters, therefore our discussion will be split into two parts: $U_{g}>2V_{g}$ and $U_{g}<2V_{g}$. In this study, we have set our parameters as follows: $V_{g} = \alpha U_{g}$, $V_{g,e} = \alpha V'_{g,e}$, $V{e} = \alpha U_{e}$, $V'_{g,e} = 1.5U_{g}$, $U_{e} = 2U_{g}$, $t_{e} = 0.3t_{g}$, $t_{g,e} = 0.6t_{g}$. Here according to the literature, the coupling strength ratio of $\alpha$ can be either set as 0.2 \cite{PhysRevB.95.241303} or 0.7 \cite{10.1038/s41534-021-00449-4,PhysRevApplied.12.064049,PhysRevB.83.235314}, and $U=U_{g}/t_{g}$ is the main quantity parameter in the results.

\label{sec:result}

\subsection{Local entanglement at $\varepsilon _{1}=\varepsilon _{2}=\varepsilon _{3}=\varepsilon _{4}=0$}

\begin{figure}
	
	(a)\includegraphics[scale=0.75]{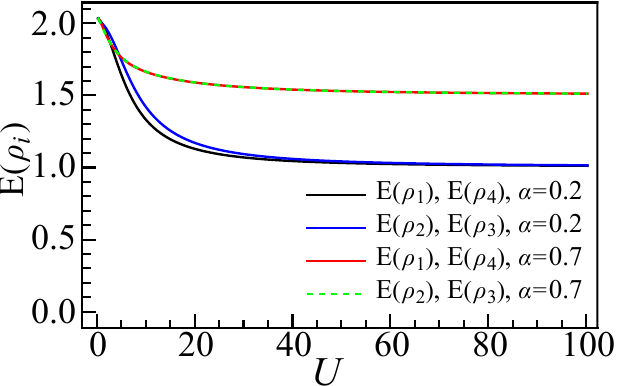}
	(b)\includegraphics[scale=0.75]{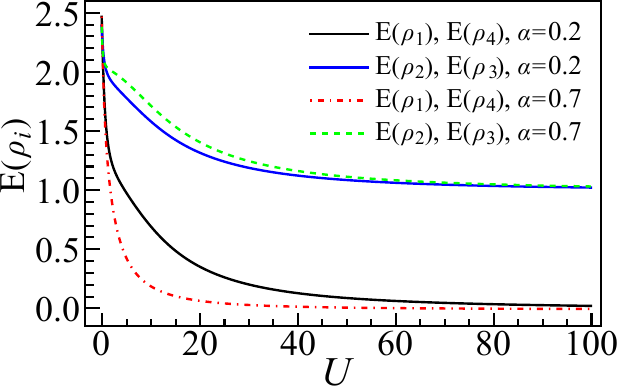}

	\caption{Local entanglement $E(\rho_{i})$ profiles for a four-site quantum dot system ($L=4$) with coupling strengths $\alpha=0.2$ or $\alpha=0.7$, displayed as a function of interaction strength $U$. Panels (a) and (b) correspond to systems with four ($N=4$) and six ($N=6$) electrons, respectively, with zero detuning energy ($\varepsilon_i=0$) at all sites. The entanglement measures $E(\rho_{1})$ and $E(\rho_{4})$ are equivalent, as are $E(\rho_{2})$ and $E(\rho_{3})$}
	\label{fig:zeroLocal}
\end{figure}

\begin{figure}
	
	(a)\includegraphics[scale=0.75]{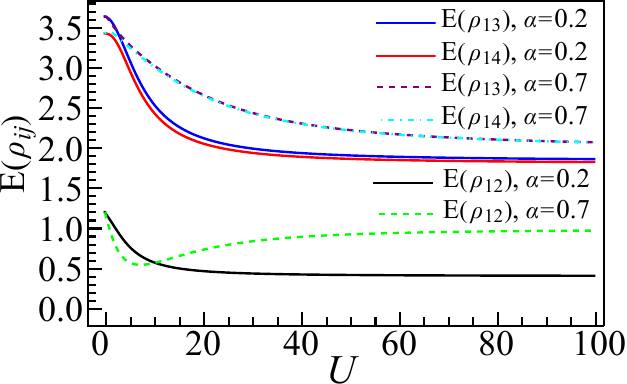}
	(b)\includegraphics[scale=0.75]{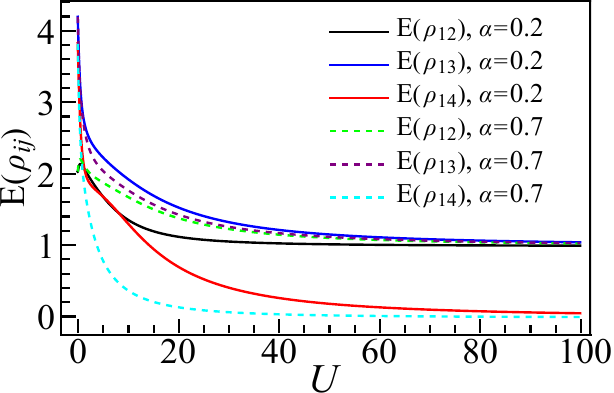}

	\caption{These figures illustrate the pairwise entanglement metrics $E(\rho_{ij})$ for a four-site ($L=4$) quantum dot array, analyzed under two coupling strength scenarios, $\alpha=0.2$ and $\alpha=0.7$. Displayed as functions of the interaction parameter $U$, panel (a) details configurations with four electrons ($N=4$), and panel (b) with six electrons ($N=6$), all with zero detuning energy at each site ($\varepsilon_i = 0$). The figures demonstrate equivalent entanglement values between dot pairs, specifically, $E(\rho_{12})$ with $E(\rho_{34})$, $E(\rho_{13})$ with $E(\rho_{24})$, and $E(\rho_{23})$ with $E(\rho_{14})$.}
	\label{fig:zeroPairwise}
\end{figure}

Starting with an analysis of the local entanglement in the smallest system size ($L=4$) for both electron number scenarios ($N=4$ and $N=6$), we first consider the case of $N=4$ with $\alpha=0.2$, as depicted in Fig.~\ref{fig:zeroLocal}(a). It is apparent that for $N=4$, the local entanglement at the end sites ($E(\rho_{1})=E(\rho_{L})$) is both equal and less than the local entanglement of the inner sites. This phenomenon arises from the preference of the end sites for single occupancy over the middle sites, particularly as the repulsive interaction increases \cite{PhysRevA.106.022421}. With the increase in the repulsive interaction $U$ in the four-dot-four-electron system, specific configurations, such as $|\uparrow_{g},\downarrow_{g},\uparrow_{g},\downarrow_{g}\rangle$, $|\downarrow_{g},\uparrow_{g},\downarrow_{g},\uparrow_{g}\rangle$, $|\uparrow_{g},\uparrow_{g},\downarrow_{g},\downarrow_{g}\rangle$, $|\downarrow_{g},\downarrow_{g},\uparrow_{g},\downarrow_{g}\rangle$, $|\uparrow_{g},\downarrow_{g},\downarrow_{g},\uparrow_{g}\rangle$, and $|\downarrow_{g},\uparrow_{g},\uparrow_{g},\downarrow_{g}\rangle$, progressively dominate the ground state, as illustrated in Fig.~\ref{fig:4ele4U0p2Vdot}(i).

For $N=4$ and $\alpha=0.7$, with $U_{g}<2V_{g}$, akin to the behavior observed in charge density wave in large chain systems \cite{PhysRevLett.93.086402}, electrons in a single dot tend to favor double occupancy over single occupancy. In a four-dot system, as $U$ increases, specific electron configurations such as $|\uparrow_{g},\downarrow_{g},0,\uparrow_{g}\downarrow_{g}\rangle$, $|\downarrow_{g},\uparrow_{g},0,\uparrow_{g}\downarrow_{g}\rangle$, $|\uparrow_{g}\downarrow_{g},0,\uparrow_{g},\downarrow_{g}\rangle$, and $|\uparrow_{g}\downarrow_{g},0,\downarrow_{g},\uparrow_{g}\rangle$ come to dominate the ground state configuration, as depicted in Fig.~\ref{fig:4ele4U0p7Vdot}(i) (the above four states are all represented by $|\uparrow_{g}\downarrow_{g},0,\uparrow_{g},\downarrow_{g}\rangle$ since they can be equally treated). This is related to the small size effect, since in such a system these configurations are most energetically favorable. Also, in Fig.~\ref{fig:zeroLocal}(a), it is evident that $E(\rho_{1})=E(\rho_{4})$ and $E(\rho_{2})=E(\rho_{3})$, as all sites have an equal ratio of the four configurations of $|0\rangle$, $|\uparrow_{g}\rangle$, $|\downarrow_{g}\rangle$, and $|\uparrow_{g}\downarrow_{g}\rangle$. Specifically, $E(\rho_{1})$ is almost equal to $E(\rho_{2})$, with any differences being brought about by configuration states such as $|\uparrow_{g}\downarrow_{g},0,0,\uparrow_{g}\downarrow_{g}\rangle$, illustrated in Fig.~\ref{fig:4ele4U0p7Vdot}(h). 

In the $L=4$, $N=6$, and $\alpha=0.2$ system, entanglement is shown in Fig.~\ref{fig:zeroLocal}(b). Due to the presence of two extra electrons (compared to the $N=4$ case), the electron configurations of $|\uparrow_{g}\downarrow_{g},\uparrow_{g},\downarrow_{g},\uparrow_{g}\downarrow_{g}\rangle$, $|\uparrow_{g}\downarrow_{g},\uparrow_{g},\uparrow_{g}\downarrow_{g},\downarrow_{g}\rangle$, and $|\uparrow_{g}\downarrow_{g},\uparrow_{g}\downarrow_{g},\uparrow_{g},\downarrow_{g}\rangle$ have the primary contribution to the system ground state $\psi_\mathrm{GS}$, as shown in Fig.~\ref{fig:4ele6U0p2Vdot}(i). For the ease of later discussion, we also introduce a notation describing the number of electrons in different sites. For example,  $|{\bullet\bullet},\bullet,\bullet,{\bullet\bullet}\rangle$, $|{\bullet\bullet},\bullet,{\bullet\bullet},\bullet\rangle$, and $|{\bullet\bullet},{\bullet\bullet},\bullet,\bullet\rangle$ represent the three aforementioned states occupancy respectively, where $\bullet$ or ${\bullet\bullet}$ represents a site occupied by one electron or two electrons respectively. We also use $\circ$ to express an empty site, so 
$|{\bullet\bullet},\circ,{\bullet\bullet},{\bullet\bullet}\rangle$ represents a case where site-1, site-3 and site-4 are doubly occupied while site-2 has no electron.

In the weak coupling regime, where $U \sim 0$, all electron configuration components have roughly the same proportion, thus $E(\rho_{i})$ at $U \sim 0$ have similar values. As $U$ increases, the local entanglement of the end dots decreases more rapidly than that of the inner dots from the middle of the chain, and this rate of descent is even faster in the $N=6$ case than the $N=4$ case with $\alpha=0.2$. This is due to the increasing dominance of the $|{\bullet\bullet},\bullet,\bullet,{\bullet\bullet}\rangle$ configuration in the ground state, as depicted in Fig.~\ref{fig:4ele6U0p2Vdot}(i). At $U\gg 1$, the inner dots tend to favor single occupancy, thereby resulting in similar values for $E(\rho_{2})$ and $E(\rho_{3})$ for both $N=4$ and $N=6$, while the end dots in the $N=6$ case favor double occupancy, leading to a rapid decrease in the entanglement value.

For $N=6$ and $\alpha=0.7$, the system tends to favor double occupancy. Hence, the configurations $|{\bullet\bullet},\bullet,\bullet,{\bullet\bullet}\rangle$, $|{\bullet\bullet},\circ,{\bullet\bullet},{\bullet\bullet}\rangle$ (also $|{\bullet\bullet},{\bullet\bullet},\circ,{\bullet\bullet}\rangle$) have a greater presence in the ground state compared to the $\alpha=0.2$ case, as illustrated in Fig.~\ref{fig:4ele6U0p7Vdot}(i). When compared to Fig.~\ref{fig:4ele6U0p2Vdot}(i), the maximal probability of $|{\bullet\bullet},\circ,{\bullet\bullet},{\bullet\bullet}\rangle$ and $|{\bullet\bullet},{\bullet\bullet},\circ,{\bullet\bullet}\rangle$ in Fig.~\ref{fig:4ele6U0p7Vdot}(i) has shifted toward smaller $U$. This indicates that all four sites in the $\alpha=0.7$ setup prefer double occupancy over the $\alpha=0.2$ case, leading to a faster decrease in $E(\rho_{1})$ and $E(\rho_{4})$, and a slower decrease in $E(\rho_{2})$ and $E(\rho_{3})$ compared to the $\alpha=0.2$ scenario, since double occupancy contributes more to local entanglement.

\subsection{Pairwise entanglement  at $\varepsilon _{1}=\varepsilon _{2}=\varepsilon _{3}=\varepsilon _{4}=0$}

In $L=4$ system with all quantum dots having equal potential energy ($\varepsilon_1 = \varepsilon_2 = \varepsilon_3 = \varepsilon_4 = 0$), mirror reflection symmetry ensures that the pairs of two-site reduced density matrices satisfy the relations $\rho_{12} = \rho_{34}$ and $\rho_{13} = \rho_{24}$. Additionally, due to the finite size effect inherent in the small system, it is observed that $\rho_{14} = \rho_{23}$, as illustrated in Figure~\ref{fig:zeroPairwise}.

For $N=4$ and $\alpha=0.2$, the entanglement results of $E(\rho_{12})$, $E(\rho_{13})$, and $E(\rho_{14})$  align well with the theoretical predictions for non-interacting systems ($\alpha=0$), as elucidated in Ref.~\cite{PhysRevA.106.022421} and depicted in Figure~\ref{fig:zeroPairwise}(a). In the limit where $U \sim 0$, $E(\rho_{ij})$ has the same value for different $\alpha$ values since all Coulomb interactions are zero. Conversely, at $\alpha=0.7$ with a positive $U$ value, the system demonstrates a preference for electron configurations such as $|{\bullet\bullet},\circ,\bullet,\bullet\rangle$ and $|\bullet,\bullet,\circ,{\bullet\bullet}\rangle$. This preference equilibrates the entanglement levels $E(\rho_{13})$ and $E(\rho_{14})$ within the strong coupling regime, as illustrated in Figure~\ref{fig:zeroPairwise}(a). Within this regime, the probabilities for zero and single electron occupancy at sites 2 and 3 become comparable, as do the probabilities for single and double electron occupancy at sites 1 and 4, a phenomenon detailed in Figure~\ref{fig:4ele4U0p7Vdot}(i). Concerning $E(\rho_{12})$, as depicted in the same figure, the diminished favorability of the state $|{\bullet\bullet},\circ\rangle$ for the first and second sites leads to a reduction in the prevalence of the state $|{\bullet\bullet},\circ,\circ,{\bullet\bullet}\rangle$ as $U$ increases. This reduction also leads to an increase in $E(\rho_{12})$ around $U \approx 7$, beyond which $E(\rho_{12})$ stabilizes to a constant value as $U$ continues to increase.

For $N=6$ and $U=0$, the uneven distribution of electrons leads to increased entanglement $E(\rho_{ij})$ compared to $N=4$. This is particularly evident for $E(\rho_{12})$, as sites 1 and 2 are more likely to adopt the $|{\bullet\bullet},\bullet\rangle$ configuration instead of the local half-filled state. Figures~\ref{fig:4ele6U0p2Vdot}(i) and~\ref{fig:4ele6U0p7Vdot}(i) illustrate that the electron arrangements $|{\bullet\bullet},\bullet,\bullet,{\bullet\bullet}\rangle$, $|{\bullet\bullet},\circ,{\bullet\bullet},{\bullet\bullet}\rangle$, and $|\bullet,{\bullet\bullet},\bullet,{\bullet\bullet}\rangle$ play a key role in determining the entanglement. For $\alpha=0.7$, double occupancy is preferred, leading to a more rapid decline in configurations like $|\bullet,{\bullet\bullet},\bullet,{\bullet\bullet}\rangle$ as $U$ increases, which in turn causes a quicker reduction in entanglement $E(\rho_{ij})$ compared to $\alpha=0.2$. Regarding $E(\rho_{14})$, as $U$ moves into the strong coupling regime, $E(\rho_{14})$ approaches zero since sites 1 and 4 predominantly favor the $|{\bullet\bullet}\rangle$ configuration.

\subsection{Entanglement analysis for $\varepsilon_1\neq 0 $ with $N=4$}

\begin{figure*}[t]\centering
	(a)\includegraphics[scale=0.42]{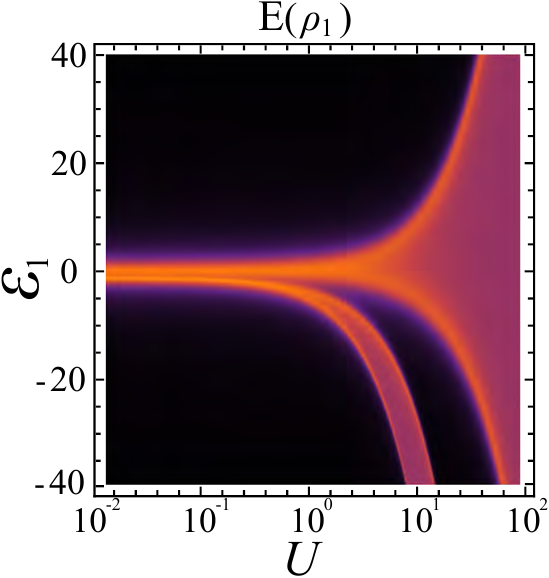}
	(b)\includegraphics[scale=0.42]{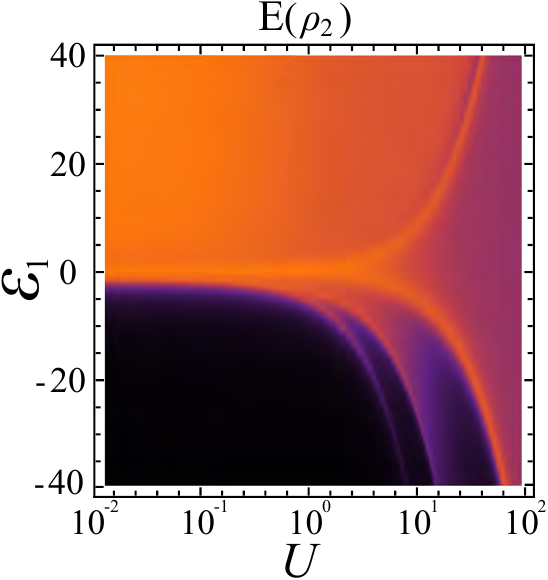}
	(c)\includegraphics[scale=0.42]{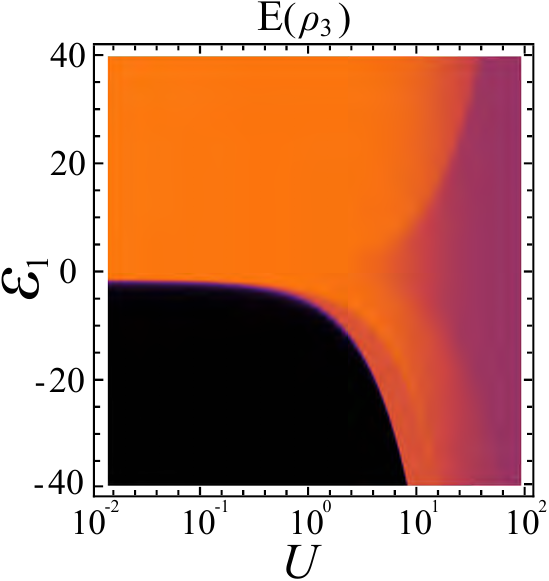}
	(d)\includegraphics[scale=0.42]{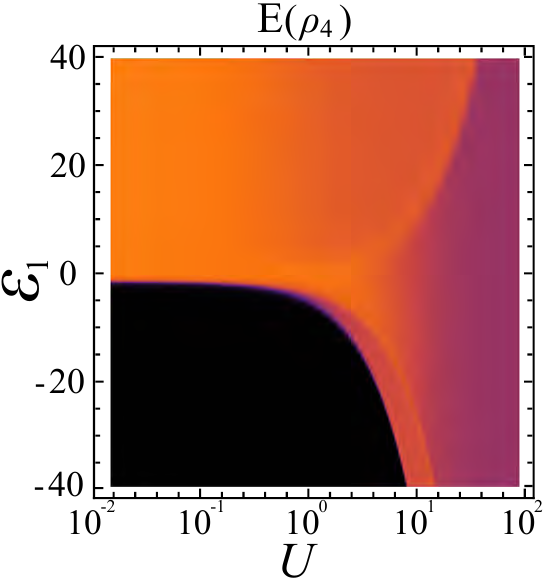}
	(e)\includegraphics[scale=0.42]{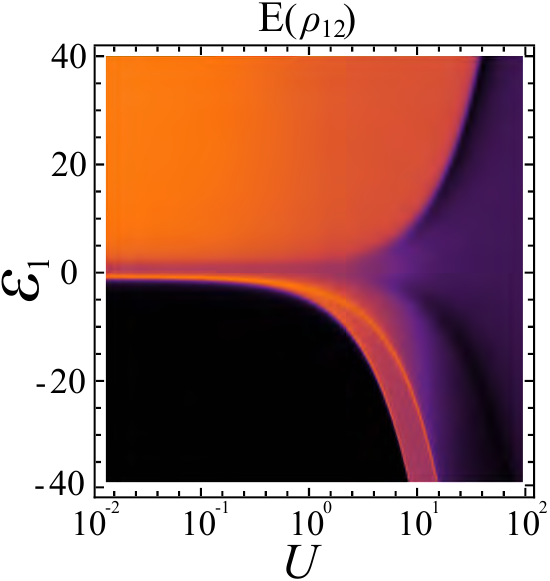}
	(f)\includegraphics[scale=0.42]{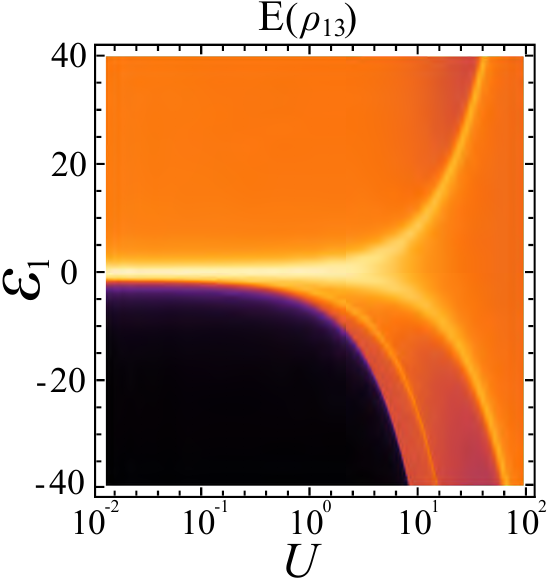}
	(g)\includegraphics[scale=0.42]{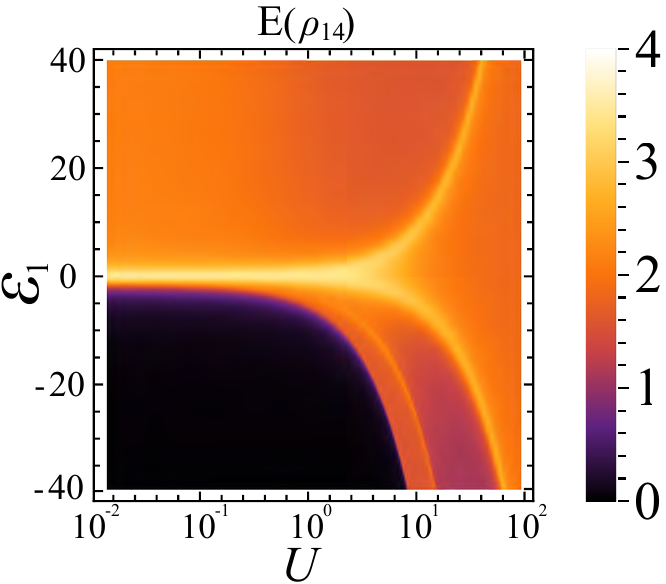}
	(h)\includegraphics[scale=0.5]{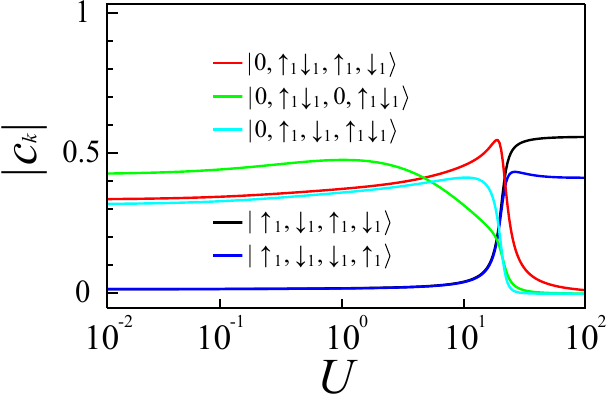}
	(i)\includegraphics[scale=0.5]{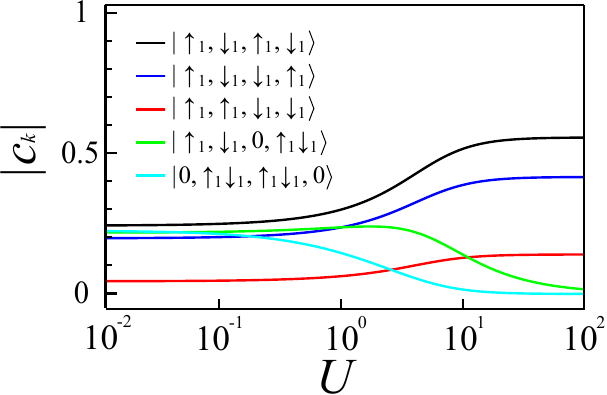}
	(j)\includegraphics[scale=0.5]{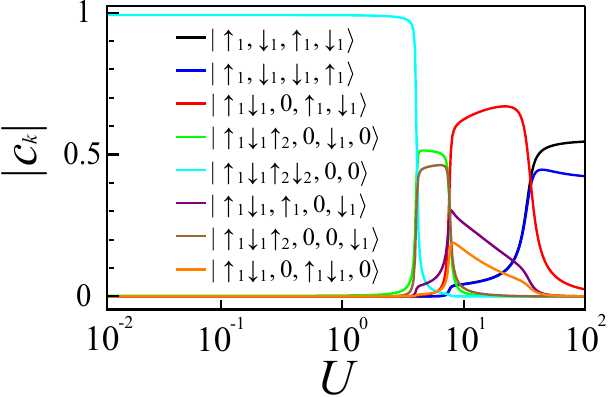}
	\caption{Entanglement phase diagrams for a quantum dot system with four sites ($L=4$) and four electrons ($N=4$) under a coupling strength ratio of $\alpha=0.2$. These diagrams are plotted as functions of the interaction strength $U$ and the potential energy $\varepsilon_{1}$. (a)-(d) local entanglement measures $E(\rho_{1})$, $E(\rho_{2})$, $E(\rho_{3})$, and $E(\rho_{4})$, respectively. (e)-(g) pairwise entanglement for dot pairs $E(\rho_{12})$, $E(\rho_{13})$, and $E(\rho_{14})$. (h)-(j) illustrate the proportions of selected advantageous electron configurations within the system's ground state, highlighting the influence of interaction parameters on system behavior, represent cases where $\varepsilon_1=20$, $\varepsilon_1=0$, and $\varepsilon_1=-20$, respectively.}\label{fig:4ele4U0p2Vdot}
\end{figure*}

\begin{figure*}[t]\centering
	(a)\includegraphics[scale=0.42]{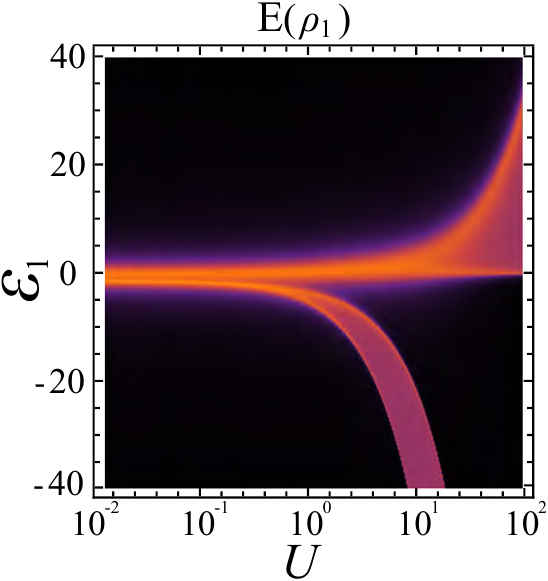}
	(b)\includegraphics[scale=0.42]{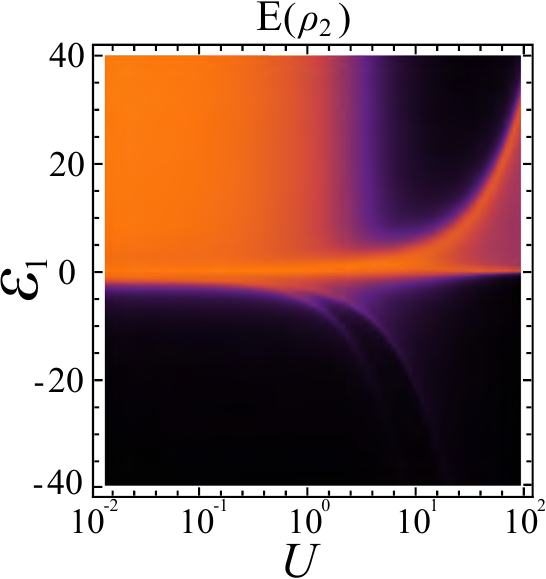}
	(c)\includegraphics[scale=0.42]{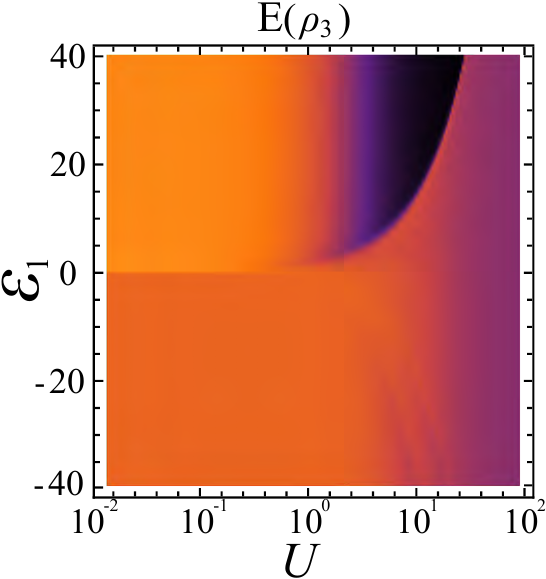}
	(d)\includegraphics[scale=0.42]{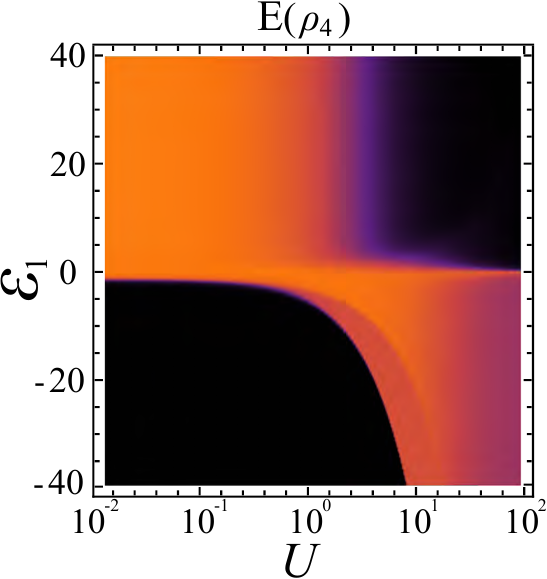}
	(e)\includegraphics[scale=0.42]{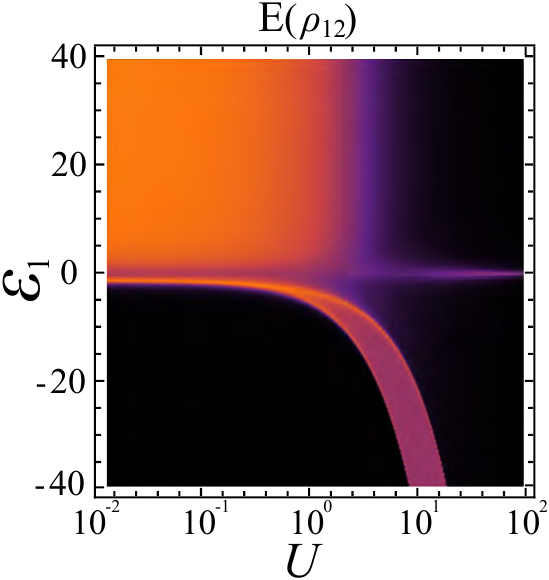}
	(f)\includegraphics[scale=0.42]{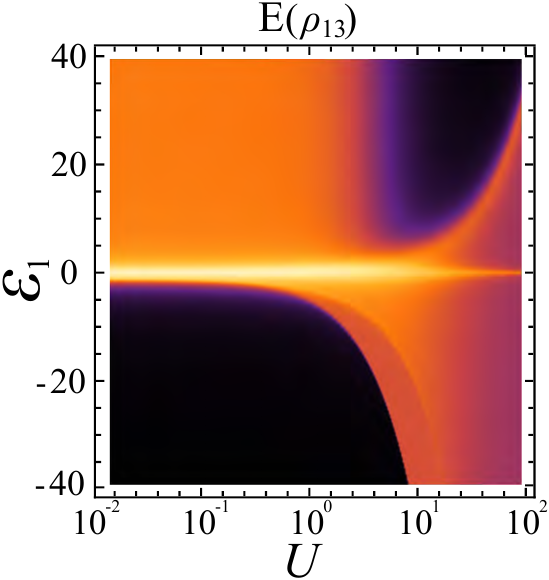}
	(g)\includegraphics[scale=0.42]{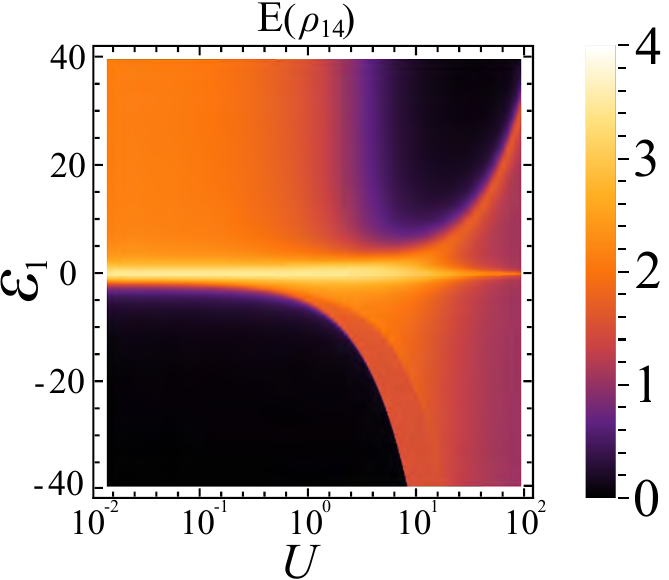}
	(h)\includegraphics[scale=0.5]{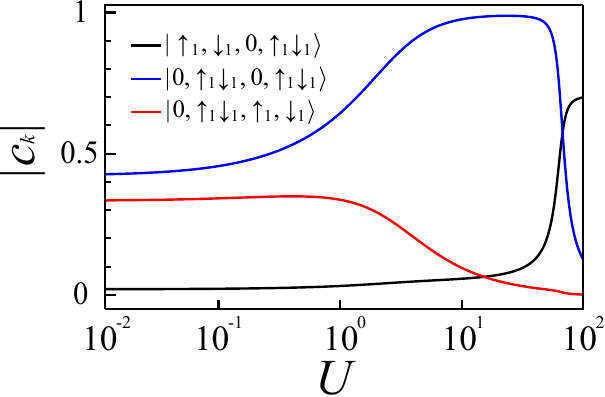}
	(i)\includegraphics[scale=0.5]{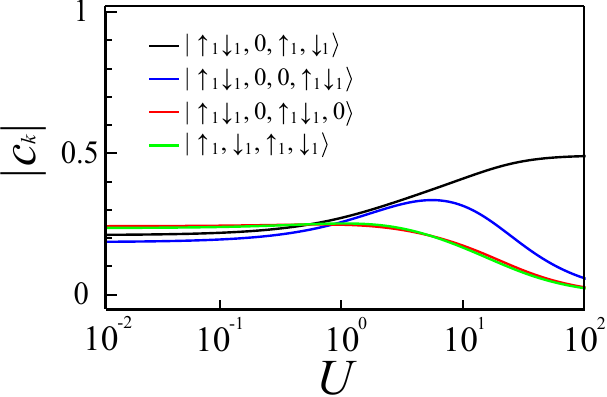}
	(j)\includegraphics[scale=0.5]{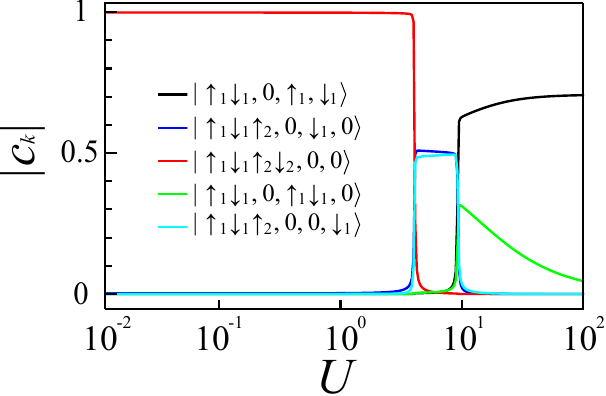}
	\caption{Entanglement characteristics of a four-dot ($L=4$), four-electron ($N=4$) quantum dot system at a coupling ratio of $\alpha=0.7$. Diagrams are plotted against interaction strength $U$ and potential energy $\varepsilon_{1}$. (a)-(d) local entanglement measures $E(\rho_{1})$ to $E(\rho_{4})$. (e)-(g) pairwise entanglement for dot pairs $E(\rho_{12})$, $E(\rho_{13})$, and $E(\rho_{14})$. The dominant electron configurations in the ground state corresponding to $\varepsilon_1 = 20$, $\varepsilon_1 = 0$, and $\varepsilon_1 = -20$ are represented by (h), (i), and (j) respectively.}\label{fig:4ele4U0p7Vdot}
\end{figure*}

\begin{figure*}[t]\centering
	(a)\includegraphics[scale=0.42]{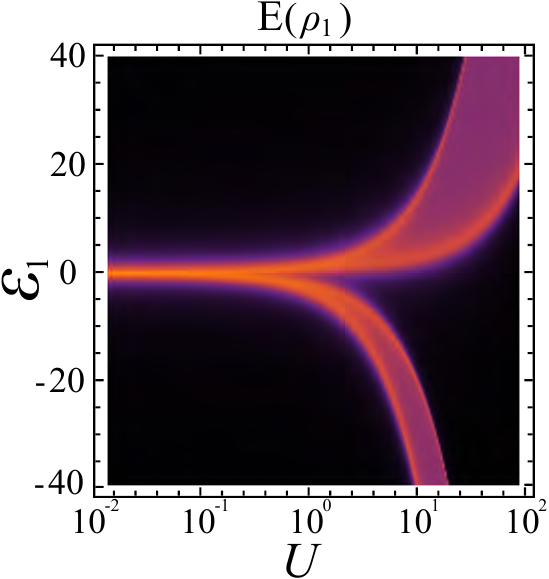}
	(b)\includegraphics[scale=0.42]{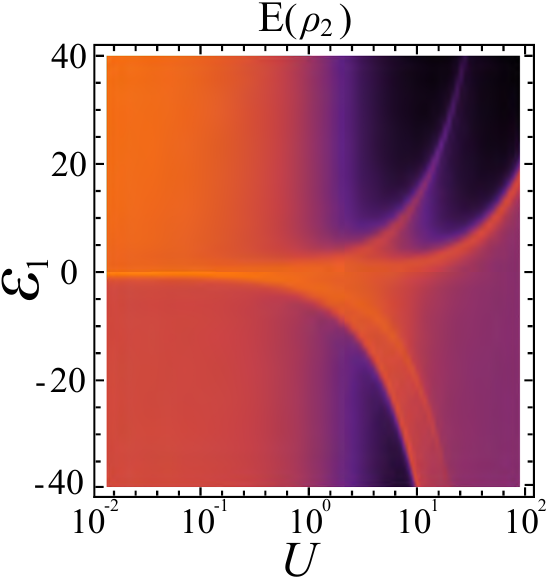}
	(c)\includegraphics[scale=0.42]{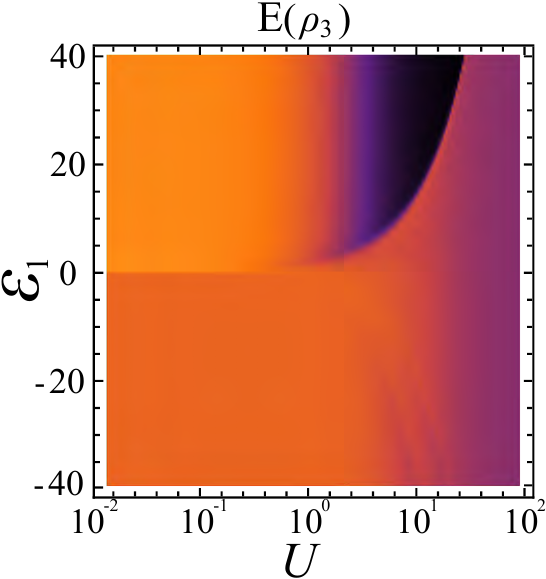}
	(d)\includegraphics[scale=0.42]{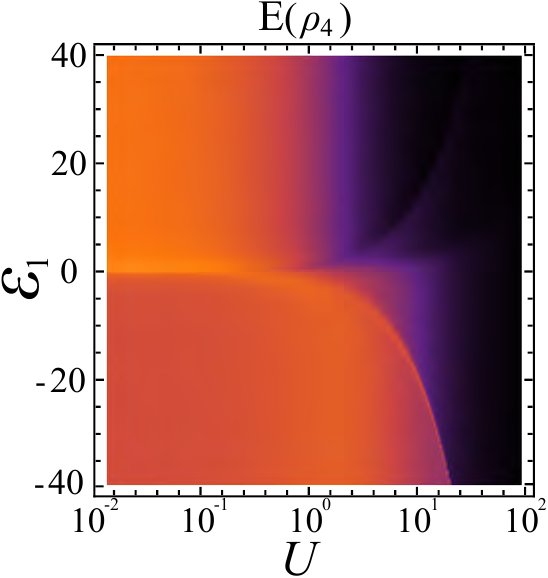}
	(e)\includegraphics[scale=0.42]{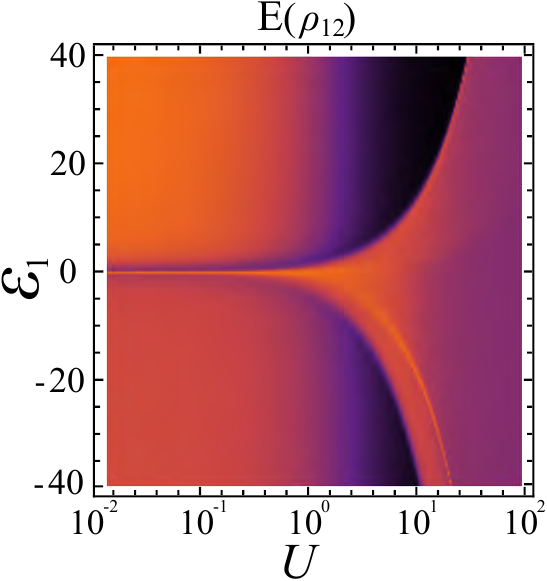}
	(f)\includegraphics[scale=0.42]{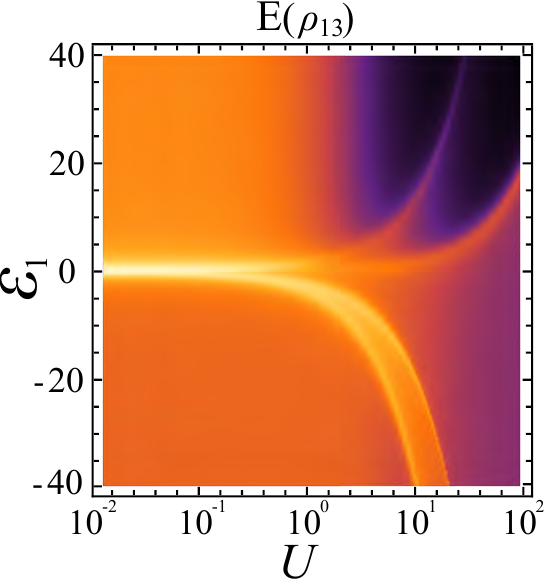}
	(g)\includegraphics[scale=0.42]{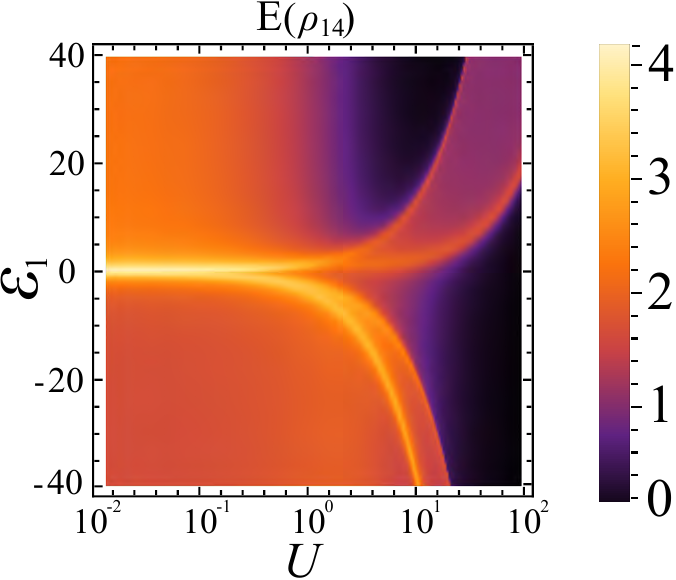}
	(h)\includegraphics[scale=0.5]{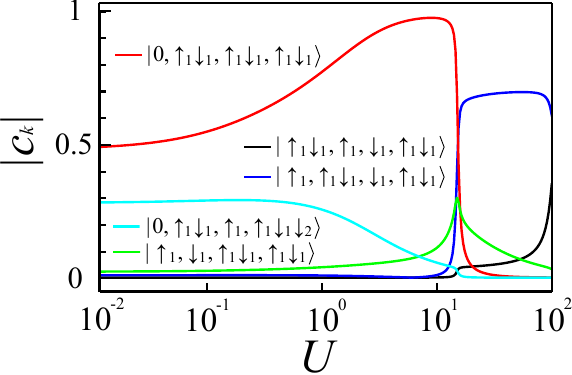}
	(i)\includegraphics[scale=0.5]{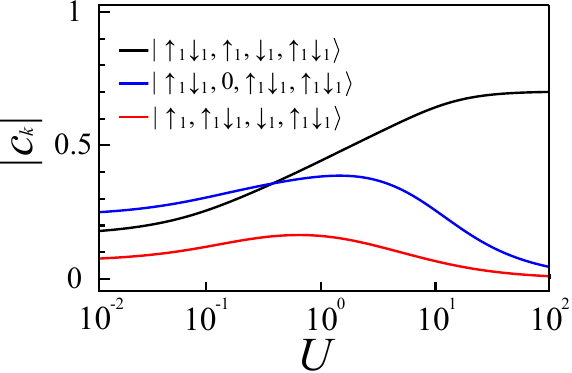}
	(j)\includegraphics[scale=0.5]{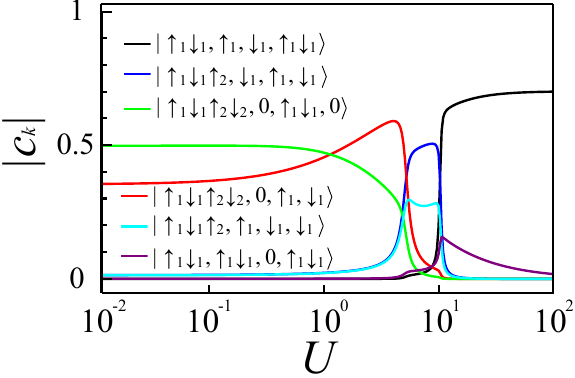}
	\caption{Entanglement profiles for a four-site ($L=4$), six-electron ($N=6$) quantum dot system with a coupling strength of $\alpha=0.2$. Charts are graphed according to interaction strength $U$ and potential energy $\varepsilon_{1}$. (a)-(d) depict local entanglement levels $E(\rho_{1})$ through $E(\rho_{4})$. (e)-(g) pairwise entanglement between dot pairs $E(\rho_{12})$, $E(\rho_{13})$, and $E(\rho_{14})$. (h)-(j) show the predominant electron configurations in the system's ground state, corresponding to scenarios where $\varepsilon_1=20$, $\varepsilon_1=0$, and $\varepsilon_1=-20$, respectively.}\label{fig:4ele6U0p2Vdot}
\end{figure*}

\begin{figure*}[t]\centering
	(a)\includegraphics[scale=0.42]{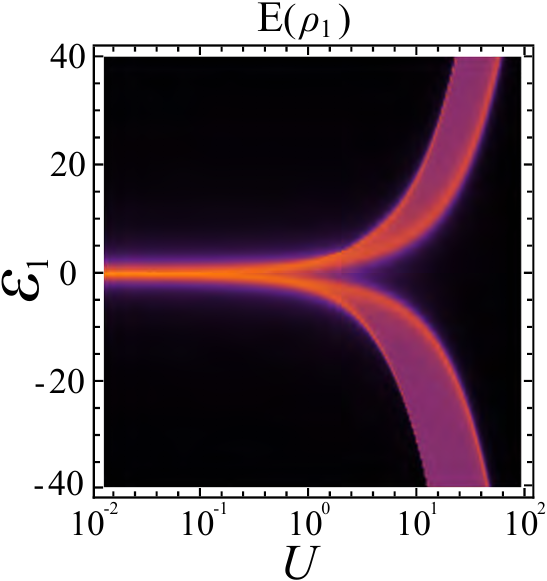}
	(b)\includegraphics[scale=0.42]{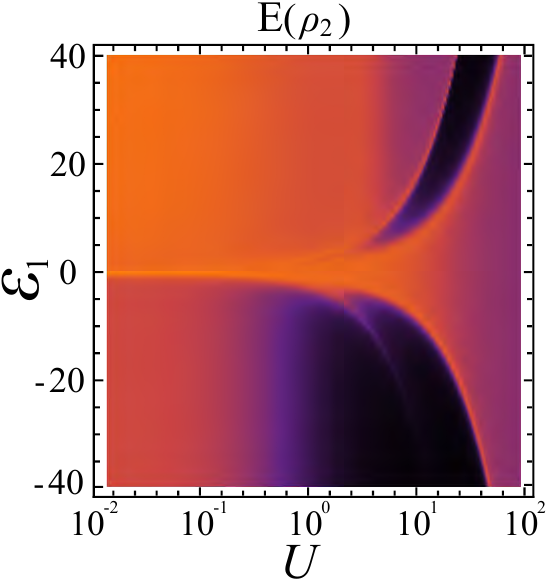}
	(c)\includegraphics[scale=0.42]{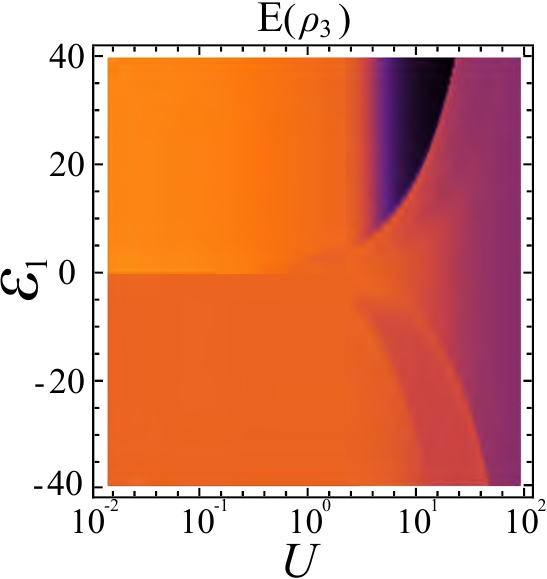}
	(d)\includegraphics[scale=0.42]{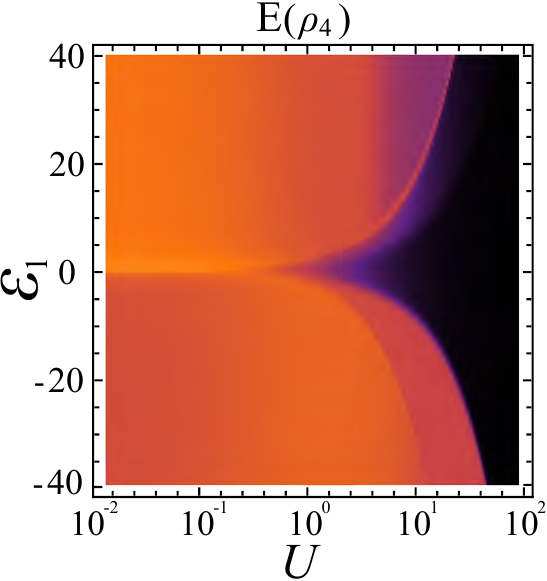}
	(e)\includegraphics[scale=0.42]{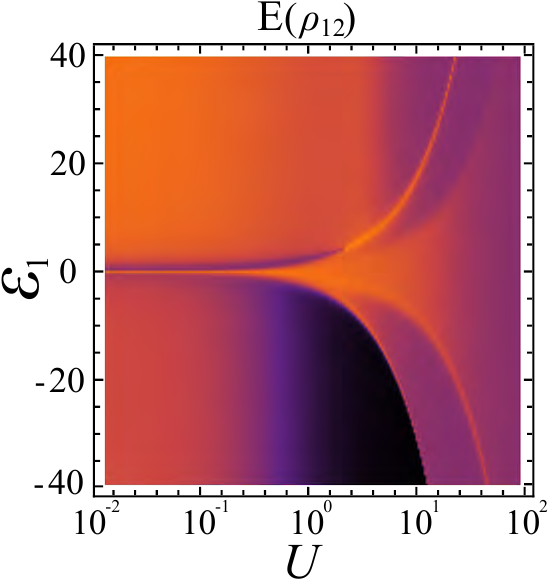}
	(f)\includegraphics[scale=0.42]{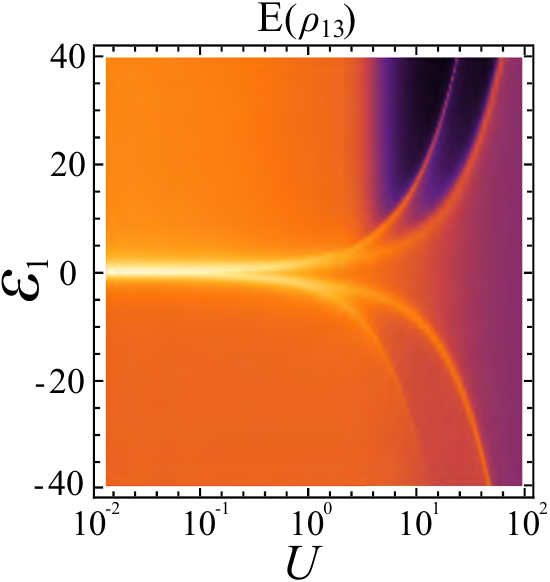}
	(g)\includegraphics[scale=0.42]{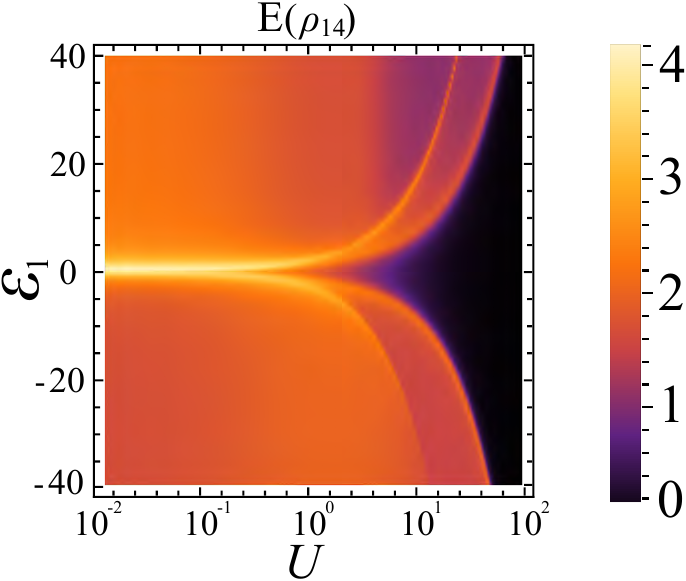}
	(h)\includegraphics[scale=0.5]{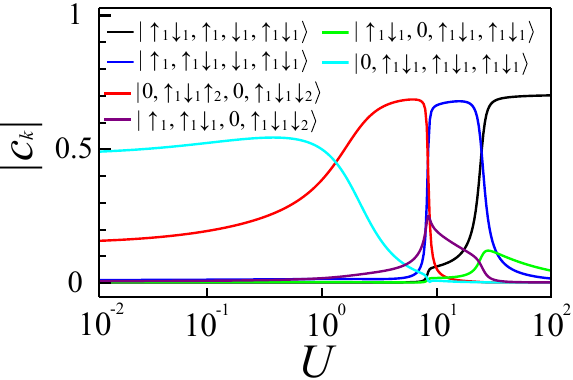}
	(i)\includegraphics[scale=0.5]{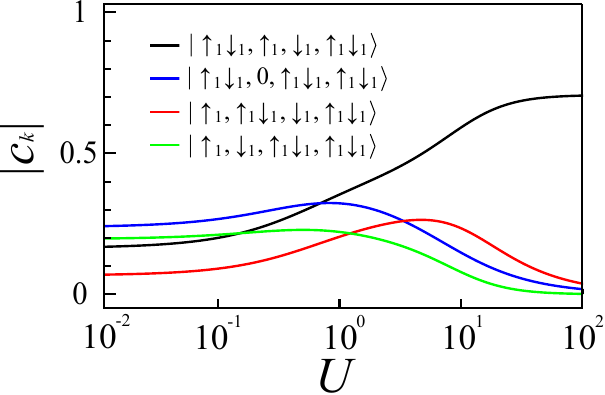}
	(j)\includegraphics[scale=0.5]{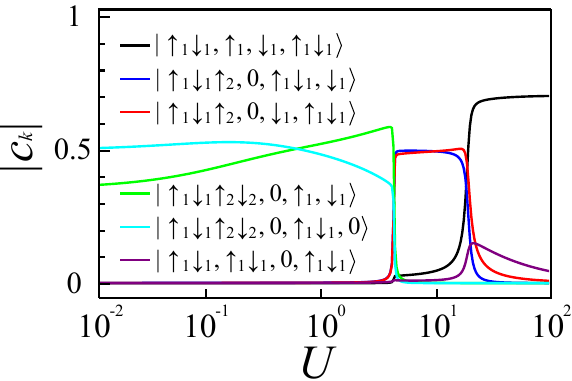}
	\caption{Entanglement profiles for a four-site ($L=4$), six-electron ($N=6$) quantum dot system with a coupling strength of $\alpha=0.7$. Charts are graphed according to interaction strength $U$ and potential energy $\varepsilon_{1}$. (a)-(d) depict local entanglement levels $E(\rho_{1})$ through $E(\rho_{4})$. (e)-(g) pairwise entanglement between dot pairs $E(\rho_{12})$, $E(\rho_{13})$, and $E(\rho_{14})$. (h)-(j) illustrate the dominant electron configurations in the system's ground state for $\varepsilon_1=20$, $\varepsilon_1=0$, and $\varepsilon_1=-20$, respectively.}\label{fig:4ele6U0p7Vdot}
\end{figure*}
Altering the potential energy of a specific quantum dot can significantly impact the entanglement behavior in the system, as demonstrated in Figures~\ref{fig:4ele4U0p2Vdot}, \ref{fig:4ele4U0p7Vdot}, \ref{fig:4ele6U0p2Vdot}, and \ref{fig:4ele6U0p7Vdot}. For a particular quantum dot $i$, decreasing its potential energy causes electrons to congregate in this dot, which is reflected in the changes in the reduced density matrix elements: $v_{i,7}$, $v_{i,8}$, and $v_{i,9}$ increase, while $v_{i,1}$ to $v_{i,6}$ decrease.

In contrast, increasing the potential energy of dot $i$ leads to the dispersal of electrons to other dots, resulting in a decrease in all matrix elements of $\rho_{i}$ except for $v_{i,1}$, which corresponds to zero electron occupancy. In extreme cases, where the potential energy $\varepsilon_{i}$ undergoes significant changes, the electron configuration in this dot transitions to either $|0\rangle$ or $|\uparrow_{g}\downarrow_{g}, \uparrow_{e}\downarrow_{e}\rangle$, causing the local entanglement value to drop to zero, as shown in Figures~\ref{fig:4ele4U0p2Vdot}(a), \ref{fig:4ele4U0p7Vdot}(a), \ref{fig:4ele6U0p2Vdot}(a), and \ref{fig:4ele6U0p7Vdot}(a). This phenomenon is particularly pronounced in the weakly coupling regime, where electrons have greater mobility. For instance, Figure~\ref{fig:4ele4U0p2Vdot}(a) depicts the relationship between local entanglement $E(\rho_{1})$, potential energy $\varepsilon_{1}$, and interaction strength $U$. In the weakly coupled regime ($U<1$), as $\varepsilon_{1}$ deviates from zero, the value of $E(\rho_{1})$ rapidly decreases from approximately 2 to 0. While in the strongly coupled regime ($U>30$), electrons tend to remain separated in their respective quantum dots, adopting spin-wave-like configurations. Consequently, the local entanglement value approaches a limit of 1 as $U$ increases.

For coupling strength ratio set as $\alpha=0.2$ and total electron number $N=4$, we examine the system's favorable occupancy configurations to understand its entanglement diagram behavior. In the regime where the potential energy $\varepsilon_{1}$ is positive, an increase in $\varepsilon_{1}$ at a constant $U$ induces a transition in the main electron occupancy configuration components of the system's ground states from mostly $|\bullet,\bullet,\bullet,\bullet\rangle$ to the collection of $|\circ,{\bullet\bullet},\circ,{\bullet\bullet}\rangle$, $|\circ,{\bullet\bullet},\bullet,\bullet\rangle$, and $|\circ,{\bullet\bullet},\bullet,\bullet\rangle$. Consequently, in the weakly coupled regime ($U<1$), $E(\rho_{1})$ undergoes a rapid decline, exhibiting distinct boundaries, while $E(\rho_{2})$, $E(\rho_{3})$, and $E(\rho_{4})$ remain largely unchanged, as depicted in Figures~\ref{fig:4ele4U0p2Vdot}(a)-(d). It is noteworthy that although the preferred electron occupancy configuration for dot 2 is $|{\bullet\bullet}\rangle$, the influence of other occupancy configurations like $|\bullet\rangle$ is significant, as shown in Figure~\ref{fig:4ele4U0p2Vdot}(h), leading to a blurred boundary in $E(\rho_{2})$.

In the strongly coupled regime ($U \gg 1$ and $\varepsilon_{1} \ll U$), the system continues to favor the $|\bullet,\bullet,\bullet,\bullet\rangle$ occupancy configuration, where a substantial potential difference is required to alter the electron configurations. This transition is depicted in Figures~\ref{fig:4ele4U0p2Vdot}(a), where an orange belt precedes the red entropy area at $\varepsilon_{1}>0$. It results from a rapid shift in preferred electron configurations, as shown in Figure~\ref{fig:4ele4U0p2Vdot}(h). Adjacent to this belt, three regimes can be distinguished based on the coupling strength and the extent of potential energy influence: (1) the potential energy-influenced weak coupling regime, where $U \sim 1$ and $\varepsilon_{1} \sim U$, allows electrons to be easily influenced by the potential energy difference between dots; (2) the potential energy-influenced strong coupling regime, representing the transition between weak and strong coupling regimes, where the potential energy can readily shift the system's favorable configurations; and (3) the strong coupling regime unaffected by potential energy, where $U \gg 1$ and the system remains largely unchanged by the relatively minor potential energy differences. These regimes are more distinguishable in the pairwise entanglement $E(\rho_{ij})$, as depicted in Figures~\ref{fig:4ele4U0p2Vdot}(e)-(g). Near this belt (the potential energy-influenced strong coupling regime), the states $|0,\uparrow_{g}\downarrow_{g}\rangle$ are highly favored for the pair $\rho_{12}$, resulting in a low entanglement value for $E(\rho_{12})$, while $|0,\uparrow_{g}\rangle$ and $|0,\downarrow_{g}\rangle$ are preferred for the pairs $\rho_{13}$ and $\rho_{14}$, leading to high entanglement values for $E(\rho_{13})$ and $E(\rho_{14})$.

In the regime where potential energy $\varepsilon _{1}<0$, multiple entanglement belts exist since one quantum dot can contain four electrons at most. In weakly coupled regimes, the decrease of potential energy $\varepsilon _{1}$ will quickly lead all electrons localized in site-1 since there are only four electrons in four quantum dots. More specifically, due to the size effect, the system is fully localized, and all entanglement values rapidly decline to zero, shown in Fig.~\ref{fig:4ele4U0p2Vdot}(a)-(g). As the coupling strength $U$ increases, as shown in Figure~\ref{fig:4ele4U0p2Vdot}(j), the favorable electron occupancy configurations of the ground states in the spin chain undergo a series of shifts: initially from $|\frac{{\bullet\bullet}}{{\bullet\bullet}},\circ,\circ,\circ\rangle$ to $|\frac{\bullet}{{\bullet\bullet}},\circ,\bullet,\circ\rangle$ and $|\frac{\bullet}{{\bullet\bullet}},\circ,\circ,\bullet\rangle$, then to $|{\bullet\bullet},\circ,\bullet,\bullet\rangle$, and eventually to $|\bullet,\bullet,\bullet,\bullet\rangle$.  Here we use $\frac{\bullet}{{\bullet\bullet}}$ or $\frac{{\bullet\bullet}}{{\bullet\bullet}}$ represents a site occupied by three electron or four electrons respectively. 

Firstly, in the intermediate phase where the preferred electron occupancy configurations are $|\frac{\bullet}{{\bullet\bullet}},\circ,\bullet,\circ\rangle$ and $|\frac{\bullet}{{\bullet\bullet}},\circ,\circ,\bullet\rangle$, three electrons tend to reside in the first dot, while the remaining electron occupies either the third or fourth dot. Consequently, $E(\rho_{1})$, $E(\rho_{3})$, and $E(\rho_{4})$ exhibit higher local entanglement values, whereas $E(\rho_{2})$ declines to a lower value, as depicted in Figures~\ref{fig:4ele4U0p2Vdot}(a)-(d). This distribution demonstrates a transition in preferred states across the quantum dots from 1 to 4. Specifically, the value of $E(\rho_{1})$ is associated with states indicative of three-electron occupancy $|\frac{\bullet}{{\bullet\bullet}}\rangle$, $E(\rho_{2})$ corresponds to zero electron occupancy $|\circ\rangle$ (or $|0\rangle$), and $E(\rho_{3})$ and $E(\rho_{4})$ oscillate between one-electron occupancy $|\bullet\rangle$ and zero occupancy $|\circ\rangle$. Similarly, $E(\rho_{12})$ exhibits a high entanglement value, while $E(\rho_{13})$ and $E(\rho_{14})$ display even higher values.

Secondly, when the system's preferred occupancy configuration is $|{\bullet\bullet},\circ,\bullet,\bullet\rangle$, the first dot favors double occupancy, and the second dot favors zero occupancy, while the third and fourth dots favor one-electron occupancy. As a result, $E(\rho_{1})$, $E(\rho_{2})$ and $E(\rho_{12})$ approach zero, while $E(\rho_{3})$, $E(\rho_{4})$, $E(\rho_{13})$ and $E(\rho_{14})$ become similar with high entanglement value as $U$ increases.

Lastly, in the region where the system favors the $|\bullet,\bullet,\bullet,\bullet\rangle$ occupancy configuration, all entanglement behaviors align with those in the $\varepsilon_1 \neq 0$ regime as the coupling strength becomes the dominant factor. Notably, the entanglement measures $E(\rho_{3})$ and $E(\rho_{4})$ exhibit smooth boundary transitions, indicating a preference for single-electron occupancy $|\bullet\rangle$ in both the third and fourth quantum dots at this boundary. Similarly to the $\varepsilon_1 = 0$ case, the system shows a preference for the configurations $|\uparrow_{g},\downarrow_{g},\uparrow_{g},\downarrow_{g}\rangle$ and $|\downarrow_{g},\uparrow_{g},\downarrow_{g},\uparrow_{g}\rangle$ over other spin state configurations, resulting in $E(\rho_{12}) < E(\rho_{13}) \sim E(\rho_{14})$.

When the coupling strength ratio is set to $\alpha=0.7$, the system exhibits a preference for double occupancy over single occupancy, in line with the nature of  EHM \cite{PhysRevB.105.115145, PhysRevB.92.075423, PhysRevB.75.165106, PhysRevLett.93.086402}. This preference is maintained even when $\varepsilon_1\neq 0$, as demonstrated in Figures~\ref{fig:4ele4U0p7Vdot}(h) and \ref{fig:4ele4U0p7Vdot}(j). For $\varepsilon_1>0$, the favored electron occupancy configuration readily becomes $|\circ,{\bullet\bullet},\circ,{\bullet\bullet}\rangle$ until $U\gg \varepsilon_1$, resulting in $E(\rho_{i})\sim E(\rho_{ij})\sim 0$ ($i$ for all sites from 1 to 4) when $U<\varepsilon_1$. Notably, in the weak coupling regime ($U\sim1$), $E(\rho_{1})$ equals zero, while $E(\rho_{2})$, $E(\rho_{3})$, $E(\rho_{4})$, $E(\rho_{12})$, $E(\rho_{13})$ and $E(\rho_{14})$ experience a decrease in entanglement value, caused by the reduction of the electron occupancy configuration $|\circ,{\bullet\bullet},\bullet,\bullet\rangle$, as shown in Figure~\ref{fig:4ele4U0p7Vdot}(h).

For $\varepsilon_1<0$, the system similarly experiences three transitions, as illustrated in Figure~\ref{fig:4ele4U0p7Vdot}(j). With increasing $U$, the electron occupancy in site-1 changes from 4 to 2, resulting in variations in the entanglement values across all sites. $E(\rho_{1})$ remains nonzero only when the average electron number in this dot is 3, due to the presence of two favored configurations, either up or down in the excited state. $E(\rho_{2})$ is predominantly zero, as this site is typically unoccupied by electrons, except along the boundary line where transitions between different system configurations render $E(\rho_{2})$ nonzero. Regarding $E(\rho_{3})$ and $E(\rho_{4})$, their electron configurations tend to converge in the strong coupling regime, resulting in similar entanglement behaviors. For $E(\rho_{12})$, the occupancy in site-1 influences the behavior of $E(\rho_{12})$, making it similar to $E(\rho_{1})$. For $E(\rho_{13})$ and $E(\rho_{14})$, their behavior in the regime where $U \gg \varepsilon_1$ is similar to $E(\rho_{3})$ and $E(\rho_{4})$, respectively. When $U \sim \varepsilon_1$, they also exhibit distinct features similar to $E(\rho_{1})$.

\subsection{Entanglement analysis for $\varepsilon_1\neq 0 $ with $N=6$}

In contrast to the $N=4$ case, the $N=6$ system in a four-site lattice ($L=4$) inherently exhibits an imbalance in electron configurations, necessitating the consideration of additional configurations.

In the strong coupling regime, where $U\gg \varepsilon_1$, Figures~\ref{fig:4ele6U0p2Vdot}(h), \ref{fig:4ele6U0p2Vdot}(j), \ref{fig:4ele6U0p7Vdot}(h), and \ref{fig:4ele6U0p7Vdot}(j) demonstrate that for both $\varepsilon_1>0$ and $\varepsilon_1<0$, and for coupling ratios $\alpha=0.2$ and $\alpha=0.7$, the system's favored occupancy configuration is $|{\bullet\bullet},\bullet,\bullet,{\bullet\bullet}\rangle$. This occupancy configuration leads to both $E(\rho_{1})$ and $E(\rho_{4})$ becoming zero, while $E(\rho_{2})$ and $E(\rho_{3})$ share the same entanglement value of approximately 1.2. When $U\sim \varepsilon_1$ and $\varepsilon_1>0$, the most favorable occupancy configuration for both $\alpha=0.2$ and $\alpha=0.7$ is $|\bullet,{\bullet\bullet},\bullet,{\bullet\bullet}\rangle$, lead to $E(\rho_{1})\sim E(\rho_{3})$ and $E(\rho_{2})\sim E(\rho_{4})$. For $\varepsilon_1<0$, the most favorable occupancy configuration is $|\frac{\bullet}{{\bullet\bullet}},\bullet,\bullet,\bullet\rangle$ for $\alpha=0.2$, and for $\alpha=0.7$, the configurations $|\frac{\bullet}{{\bullet\bullet}},\circ,{\bullet\bullet},\bullet\rangle$ and $|\frac{\bullet}{{\bullet\bullet}},\circ,\bullet,{\bullet\bullet}\rangle$ are preferred. For $\alpha=0.2$, the $|\bullet\rangle$ configuration of site-2 lead $E(\rho_{2})$ and $E(\rho_{12})$ become non-zero, which is opposite for $\alpha=0.7$ since site-2 favor $|\circ\rangle$ configuration. Therefore $E(\rho_{2})$ becomes zero and $E(\rho_{12})$ behaves like $E(\rho_{1})$. 

In the weak coupling regime with $\varepsilon_1>0$, the system prefers specific electron configurations based on the coupling strength ratio $\alpha$. For $\alpha=0.2$, the favored configurations are $|\circ,{\bullet\bullet},{\bullet\bullet},{\bullet\bullet}\rangle$ and $|\circ,{\bullet\bullet},\bullet,\frac{\bullet}{{\bullet\bullet}}\rangle$, while for $\alpha=0.7$, the preferences shift to $|\circ,{\bullet\bullet},{\bullet\bullet},{\bullet\bullet}\rangle$ and $|\circ,\frac{\bullet}{{\bullet\bullet}},\circ,\frac{\bullet}{{\bullet\bullet}}\rangle$. As $U$ increases within this regime, a transition occurs: for $\alpha=0.2$, the system changes towards occupancy $|\circ,{\bullet\bullet},{\bullet\bullet},{\bullet\bullet}\rangle$, causing all entanglement measures $E(\rho_{i})$ and $E(\rho_{ij})$ to vanish. In contrast, for $\alpha=0.7$, the system evolves towards the configuration $|\circ,\frac{\bullet}{{\bullet\bullet}},\circ,\frac{\bullet}{{\bullet\bullet}}\rangle$, leading to a vanishing of $E(\rho_{1})$, $E(\rho_{3})$, and $E(\rho_{13})$, while $E(\rho_{2})$, $E(\rho_{4})$, $E(\rho_{12})$, and $E(\rho_{14})$ stabilize at a constant value.

In the weak coupling regime with $\varepsilon_1<0$, all four electrons are in the first dot. Both for $\alpha=0.2$ and $\alpha=0.7$, the system exhibits a preference for the configurations $|\frac{{\bullet\bullet}}{{\bullet\bullet}},\circ,\bullet,\bullet\rangle$ and $|\frac{{\bullet\bullet}}{{\bullet\bullet}},\circ,{\bullet\bullet},\circ\rangle$. As a result, in these two coupling ratio settings, the entanglement measures $E(\rho_{i})$ and $E(\rho_{ij})$ display similar patterns: $E(\rho_{1})$ remains at zero, $E(\rho_{2})$ and $E(\rho_{12})$ gently descend to zero, while $E(\rho_{3})$, $E(\rho_{4})$, $E(\rho_{13})$, and $E(\rho_{14})$ find equilibrium at a constant value. Notably, the values of $E(\rho_{i})$ and $E(\rho_{ij})$ differ between $\alpha=0.2$ and $\alpha=0.7$, caused by different electron configuration ratio.

\subsection{Boundaries of Entanglement diagrams for large system}

\begin{figure*}[t]\centering
	(a)\includegraphics[scale=0.76]{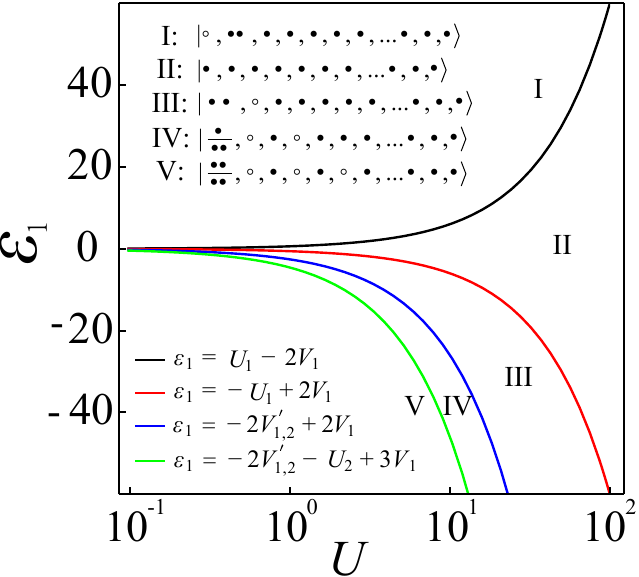}
	(b)\includegraphics[scale=0.76]{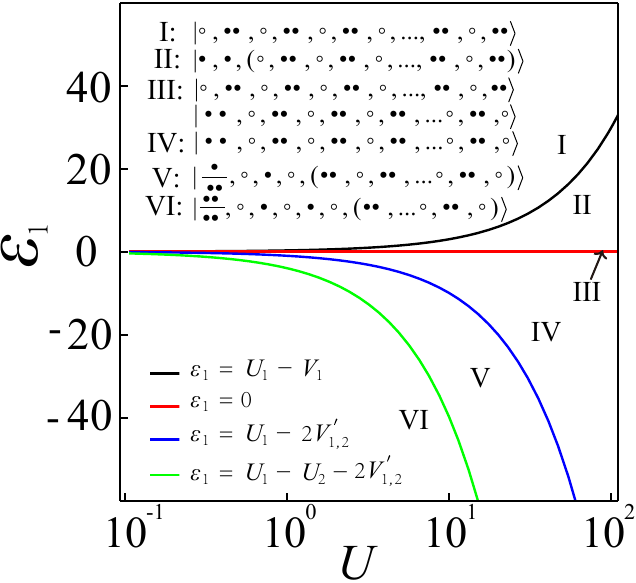}
	(c)\includegraphics[scale=0.76]{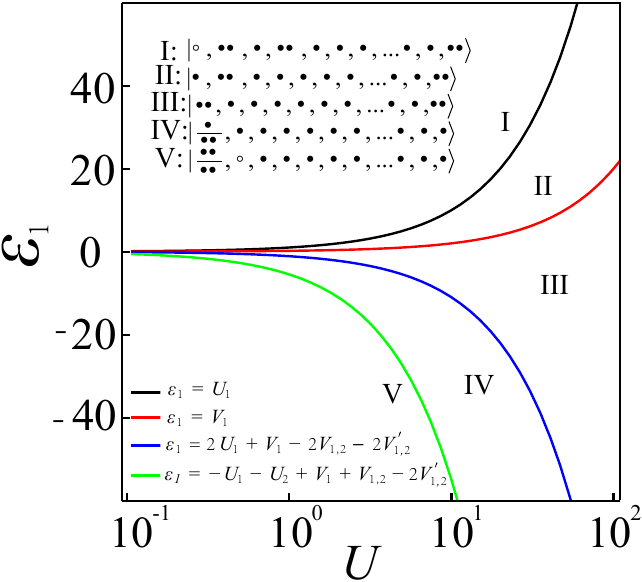}
	(d)\includegraphics[scale=0.76]{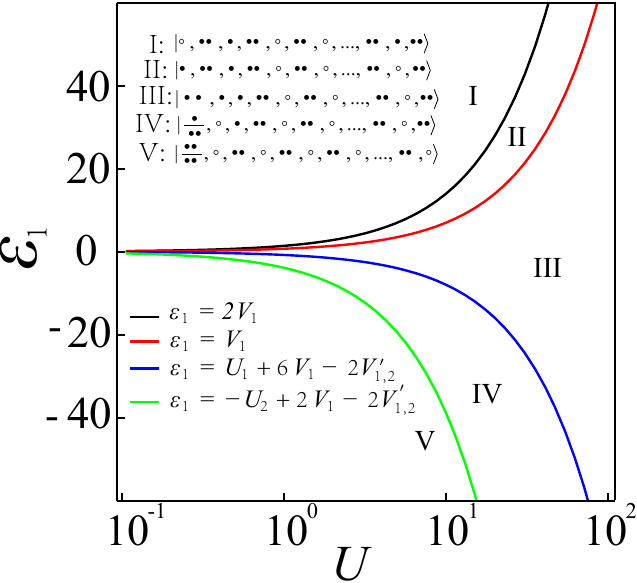}\textbf{}
	\caption{The boundaries in the entanglement diagrams for large systems, which are derived by analyzing the energy of dominant configurations as determined by the EHM. For $N=L$, (a) $\alpha=0.2$, (b) $\alpha=0.7$. For $N=L+2$, (c) $\alpha=0.2$, (d) $\alpha=0.7$.}\label{fig:phase}
\end{figure*}

In this section, we expand the entanglement diagram from a small, finite-size system to a larger spin chain quantum dot system. It is evident from the ground state of the finite-size system that advantageous electron configurations significantly influence the boundaries and values of the entanglement diagram. This analysis can be readily extended to larger systems by calculating the energy of the electron configuration obtained from the Hubbard model (see Eq.~\eqref{eq:Hubbard}). Since the system always favors the configuration with the lowest energy, which can be easily calculated and observed in small systems, we use this principle to infer the most favored configuration in larger systems.

For the case where $\alpha=0.2$ and $N=L$, with $L$ denoting the length of the spin chain and indicating an average of one electron per quantum dot, the system exhibits a preference for single occupancy at each quantum dot, resulting in a spin density wave structure as described in previous studies \cite{PhysRevB.105.115145, PhysRevB.92.075423, PhysRevB.75.165106, PhysRevLett.93.086402}.
Figure~\ref{fig:phase}(a) depicts the evolution of the dominant system configurations as the potential energy $\varepsilon_1$ transitions from positive to negative values, showcasing a sequence of dominant configurations across regimes \textbf{\uppercase\expandafter{\romannumeral1}} to \textbf{\uppercase\expandafter{\romannumeral5}}:
\begin{subequations}
	\begin{alignat}{2}
		\textbf{\uppercase\expandafter{\romannumeral1}} &: &\quad \left|\circ,{\bullet\bullet},\bullet,\bullet,\bullet,\bullet,\bullet,...\bullet,\bullet,\bullet\right\rangle, \\
		\textbf{\uppercase\expandafter{\romannumeral2}} &: &\quad \left|\bullet,\bullet,\bullet,\bullet,\bullet,\bullet,\bullet,...\bullet,\bullet,\bullet\right\rangle, \\
		\textbf{\uppercase\expandafter{\romannumeral3}} &: &\quad \left|{\bullet\bullet},\circ,\bullet,\bullet,\bullet,\bullet,\bullet,...\bullet,\bullet,\bullet\right\rangle, \\
		\textbf{\uppercase\expandafter{\romannumeral4}} &: &\quad \left|\frac{\bullet}{{\bullet\bullet}},\circ,\bullet,\circ,\bullet,\bullet,\bullet,...\bullet,\bullet,\bullet\right\rangle, \\
		\textbf{\uppercase\expandafter{\romannumeral5}} &: &\quad \left|\frac{{\bullet\bullet}}{{\bullet\bullet}},\circ,\bullet,\circ,\bullet,\circ,\bullet,...\bullet,\bullet,\bullet\right\rangle.
	\end{alignat}
\end{subequations}

The energies associated with these configurations, as derived from the Hubbard model, are as follows:
\textbf{\uppercase\expandafter{\romannumeral1}}: $U_{g}+(N-3)V_{g}$, \textbf{\uppercase\expandafter{\romannumeral2}}: $(N-1)V_{g}+\varepsilon_{1}$, \textbf{\uppercase\expandafter{\romannumeral3}}: $(N-3)V_{g}+U_{g}+2\varepsilon_{1}$, \textbf{\uppercase\expandafter{\romannumeral4}}: $(N-5)V_{g}+U_{g}+2V'_{g,e}+3\varepsilon_{1}$, \textbf{\uppercase\expandafter{\romannumeral5}}: $(N-7)V_{g}+U_{g}+U_{e}+4V'_{g,e}+4\varepsilon_{1}$. Consequently, the boundaries distinguishing these regions in Figure~\ref{fig:phase}(a) can be calculated as follows: 

\begin{subequations}
	\begin{alignat}{2}
		\textbf{\uppercase\expandafter{\romannumeral1}-\uppercase\expandafter{\romannumeral2}}: &\quad & \varepsilon_{1} &= U_{g}-2V_{g}, \\
		\textbf{\uppercase\expandafter{\romannumeral2}-\uppercase\expandafter{\romannumeral3}}: &\quad & \varepsilon_{1} &= -U_{g}+2V_{g}, \\
		\textbf{\uppercase\expandafter{\romannumeral3}-\uppercase\expandafter{\romannumeral4}}: &\quad & \varepsilon_{1} &= -2V'_{g,e}+2V_{g}, \\
		\textbf{\uppercase\expandafter{\romannumeral4}-\uppercase\expandafter{\romannumeral5}}: &\quad & \varepsilon_{1} &= -2V'_{g,e}-U_{e}+3V_{g}.
	\end{alignat}
\end{subequations}

The boundaries between different regions mark the transitions between different electron occupancy configurations. In region \textbf{I}, the system exhibits a preference for the configuration $|\circ,{\bullet\bullet},\bullet,\bullet,\bullet,\bullet,\bullet,...\bullet,\bullet,\bullet\rangle$. This indicates that the first dot is unoccupied when $\varepsilon_{1}>U_{g}-2V_{g}$, and the extra electron from the first dot is likely to be found either in the second dot or at the last dot of the spin chain. This preference arises because the electron at these positions contributes only one $V_{\nu}$ interaction, while electrons in other positions contribute to $2V_{\nu}$ interactions. Similarly, in regions \textbf{III}, \textbf{IV}, and \textbf{V}, when the first dot accommodates more than one electron, the second dot tends to be unoccupied. This arrangement minimizes the Coulomb interaction between the first and second dots. Similarly, for the remaining dots, electrons tend to favor configurations where both neighboring dots are unoccupied and form spin density wave configurations, reducing the overall Coulomb interaction terms within the system.

Second, for $\alpha=0.7$ with $N=L$, the system adopts a charge density wave structure \cite{PhysRevB.105.115145, PhysRevB.92.075423, PhysRevB.75.165106, PhysRevLett.93.086402}. Figure~\ref{fig:phase}(b) illustrates the progression of the dominant system configurations as the potential energy $\varepsilon_1$ shifts from positive to negative values, depicting a sequence of configurations that emerge in this transition. The configurations are:
\begin{subequations}
	\begin{alignat}{2}
		\textbf{\uppercase\expandafter{\romannumeral1}} &: &\quad |\circ,{\bullet\bullet},\circ,{\bullet\bullet},\circ,{\bullet\bullet},\circ,...,{\bullet\bullet},\circ,{\bullet\bullet}\rangle, \\
		\textbf{\uppercase\expandafter{\romannumeral2}} &: &\quad |\bullet, \bullet,(\circ,{\bullet\bullet},\circ,{\bullet\bullet},\circ,...,{\bullet\bullet},\circ,{\bullet\bullet})\rangle, \\
		\textbf{\uppercase\expandafter{\romannumeral3}} &: &\quad |\circ,{\bullet\bullet},\circ,{\bullet\bullet},\circ,{\bullet\bullet},\circ,...,{\bullet\bullet},\circ,{\bullet\bullet}\rangle, \\
		& &\quad |{\bullet\bullet},\circ,{\bullet\bullet},\circ,{\bullet\bullet},\circ,{\bullet\bullet},...\circ,{\bullet\bullet},\circ\rangle, \\
		\textbf{\uppercase\expandafter{\romannumeral4}} &: &\quad |({\bullet\bullet},\circ,{\bullet\bullet},\circ,{\bullet\bullet},\circ,{\bullet\bullet},...\circ,{\bullet\bullet},\circ)\rangle, \\
		\textbf{\uppercase\expandafter{\romannumeral5}} &: &\quad \left|\frac{\bullet}{{\bullet\bullet}},\circ,\bullet,\circ,({\bullet\bullet},\circ,{\bullet\bullet},...\circ,{\bullet\bullet},\circ)\right\rangle, \\
		\textbf{\uppercase\expandafter{\romannumeral6}} &: &\quad \left|\frac{{\bullet\bullet}}{{\bullet\bullet}},\circ,\bullet,\circ,\bullet,\circ,({\bullet\bullet},...\circ,{\bullet\bullet},\circ)\right\rangle.
	\end{alignat}
\end{subequations}

The energies corresponding to these configurations, as calculated from the Hubbard model, are as follows: 
\textbf{\uppercase\expandafter{\romannumeral1}}: $NU_{g}/2$, \textbf{\uppercase\expandafter{\romannumeral2}}: $(N-2)U_{g}/2+V_{g}+\varepsilon_{1}$, \textbf{\uppercase\expandafter{\romannumeral3}}: $NU_{g}/2$, \textbf{\uppercase\expandafter{\romannumeral4}}: $NU_{g}/2+2\varepsilon_{1}$, \textbf{\uppercase\expandafter{\romannumeral5}}: $(N-4)U_{g}/2+U_{g}+V'_{g,e}+3\varepsilon_{1}$
\textbf{\uppercase\expandafter{\romannumeral6}}: $(N-6)U_{g}/2+U_{g}+U_{e}+4V'_{g,e}+4\varepsilon_{1}$. Therefore, we can calculate the boundary functions of these regions in Fig.~\ref{fig:phase}(b) as 
\begin{subequations}
	\begin{alignat}{2}
		\textbf{\uppercase\expandafter{\romannumeral1}-\uppercase\expandafter{\romannumeral2}}: &\quad & \varepsilon_{1} &= U_{g}-V_{g}, \\
		\textbf{\uppercase\expandafter{\romannumeral2}-\uppercase\expandafter{\romannumeral3}}: &\quad & \varepsilon_{1} &= 0, \\
		\textbf{\uppercase\expandafter{\romannumeral3}-\uppercase\expandafter{\romannumeral4}}: &\quad & \varepsilon_{1} &= 0, \\
		\textbf{\uppercase\expandafter{\romannumeral4}-\uppercase\expandafter{\romannumeral5}}: &\quad & \varepsilon_{1} &= U_{g}-2V'_{g,e}, \\
		\textbf{\uppercase\expandafter{\romannumeral5}-\uppercase\expandafter{\romannumeral6}}: &\quad & \varepsilon_{1} &= U_{g}-U_{e}-2V'_{g,e}.
	\end{alignat}
\end{subequations}

In regions \textbf{\uppercase\expandafter{\romannumeral1}} and \textbf{\uppercase\expandafter{\romannumeral4}}, the system adopts a global charge density wave structure, with the first dot being unoccupied and doubly occupied, respectively. Notably, in region \textbf{\uppercase\expandafter{\romannumeral3}}, the system exhibits a charge density wave pattern that arises from the superposition of two distinct configurations $|\circ,{\bullet\bullet},\circ,{\bullet\bullet},\circ,{\bullet\bullet},\circ,...,{\bullet\bullet},\circ,{\bullet\bullet}\rangle$ and $|{\bullet\bullet},\circ,{\bullet\bullet},\circ,{\bullet\bullet},\circ,{\bullet\bullet},...\circ,{\bullet\bullet},\circ\rangle$. In region \textbf{\uppercase\expandafter{\romannumeral2}}, both the first and second dots host a single electron. This arrangement minimizes the Coulomb interaction terms compared to alternative configurations. Specifically, the potential energy shift in the first dot and its interaction with the second dot yield a lower energy of \(\varepsilon_1 + V_g\). In contrast, hosting two electrons in the second dot would result in a higher energy, given by \(U_g\), thus making the single-electron configuration energetically favorable. For regions \textbf{\uppercase\expandafter{\romannumeral5}} and \textbf{\uppercase\expandafter{\romannumeral6}}, apart from the first dot, the system prefers configurations where neighboring dots are unoccupied, maintaining the charge density wave structure throughout the rest of the system.

Third, in the case of $\alpha=0.2$ with $N=L+2$, the presence of two additional electrons raises the average electron count per dot above one. Consequently, only a portion of the system continues to exhibit a spin density wave structure. As illustrated in Figure~\ref{fig:phase}(c), the dominant system configurations evolve as the potential energy $\varepsilon_1$ transitions from positive to negative values. The sequence of dominant configurations for regions \textbf{\uppercase\expandafter{\romannumeral1}} to \textbf{\uppercase\expandafter{\romannumeral5}} is as follows:
\begin{subequations}
	\begin{alignat}{2}
		\textbf{\uppercase\expandafter{\romannumeral1}} &: &\quad |\circ,{\bullet\bullet},\bullet,{\bullet\bullet},\bullet,\bullet,\bullet,...\bullet,\bullet,{\bullet\bullet}\rangle, \\
		\textbf{\uppercase\expandafter{\romannumeral2}} &: &\quad |\bullet,{\bullet\bullet},\bullet,\bullet,\bullet,\bullet,\bullet,...\bullet,\bullet,{\bullet\bullet}\rangle, \\
		\textbf{\uppercase\expandafter{\romannumeral3}} &: &\quad |{\bullet\bullet},\bullet,\bullet,\bullet,\bullet,\bullet,\bullet,...\bullet,\bullet,{\bullet\bullet}\rangle, \\
		\textbf{\uppercase\expandafter{\romannumeral4}} &: &\quad \left|\frac{\bullet}{{\bullet\bullet}},\bullet,\bullet,\bullet,\bullet,\bullet,\bullet,...\bullet,\bullet,\bullet\right\rangle, \\
		\textbf{\uppercase\expandafter{\romannumeral5}} &: &\quad \left|\frac{{\bullet\bullet}}{{\bullet\bullet}},\circ,\bullet,\bullet,\bullet,\bullet,\bullet,...\bullet,\bullet,\bullet\right\rangle.
	\end{alignat}
\end{subequations}

The energy associated with each configuration in the Hubbard model obtained as follows: 
\textbf{\uppercase\expandafter{\romannumeral1}}: $3U_{g}+(N-6)V_{g}+8V_{g}$, \textbf{\uppercase\expandafter{\romannumeral2}}: $\varepsilon_{1}+2U_{g}+(N-4)V_{g}+6V_{g}$, \textbf{\uppercase\expandafter{\romannumeral3}}: $2\varepsilon_{1}+(N-3)V_{g}+2U_{g}+4V_{g}$, \textbf{\uppercase\expandafter{\romannumeral4}}: $3\varepsilon_{1}+(N-2)V_{g}+2V_{g}+V_{g,e}+2V'_{g,e}$, \textbf{\uppercase\expandafter{\romannumeral5}}: $4\varepsilon_{1}+(N-3)V_{g}+U_{g}+U_{e}+4V'_{g,e}$. Accordingly, the boundary functions distinguishing these regions in Figure~\ref{fig:phase}(c) are calculated as:
\begin{subequations}
	\begin{alignat}{2}
		\textbf{\uppercase\expandafter{\romannumeral1}-\uppercase\expandafter{\romannumeral2}}: &\quad & \varepsilon_{1} &= U_{g}, \\
		\textbf{\uppercase\expandafter{\romannumeral2}-\uppercase\expandafter{\romannumeral3}}: &\quad & \varepsilon_{1} &= V_{g}, \\
		\textbf{\uppercase\expandafter{\romannumeral3}-\uppercase\expandafter{\romannumeral4}}: &\quad & \varepsilon_{1} &= V_{g}+2U_{g}-V_{g,e}-2V'_{g,e}, \\
		\textbf{\uppercase\expandafter{\romannumeral4}-\uppercase\expandafter{\romannumeral5}}: &\quad & \varepsilon_{1} &= V_{g}+V_{g,e}-U_{g}-U_{e}-2V'_{g,e},
	\end{alignat}
\end{subequations}

In region \textbf{\uppercase\expandafter{\romannumeral1}}, the first dot is unoccupied, prompting the three additional electrons to distribute themselves  along the chain to minimize Coulomb interactions: two electrons position themselves at the ends, while the third occupies a central position. This arrangement ensures minimal interaction with the electrons at the ends. Similarly, in regions \textbf{\uppercase\expandafter{\romannumeral2}} and \textbf{\uppercase\expandafter{\romannumeral3}}, the additional electrons also preferentially reside at the chain ends. Conversely, in regions \textbf{\uppercase\expandafter{\romannumeral4}} and \textbf{\uppercase\expandafter{\romannumeral5}}, the extra electrons occupy the first dot, freeing up space along the rest of the chain for one electron per dot. Notably, in region \textbf{\uppercase\expandafter{\romannumeral4}}, the electron in the second quantum dot remains localized rather than migrating to the third dot or further along the chain. This localization is evident when considering the configuration \(\left|\frac{\bullet}{{\bullet\bullet}},\bullet,\bullet,\bullet,\bullet,\bullet,\bullet,\ldots,\bullet,\bullet,\bullet\right\rangle\), where the electron in the second dot interacts with its adjacent electron with an energy of \(3V_{g}+V_{g,e}\). Conversely, in the competitive configuration \(\left|\frac{\bullet}{{\bullet\bullet}},\circ,{\bullet\bullet},\bullet,\bullet,\bullet,\bullet,\ldots,\bullet,\bullet,\bullet\right\rangle\), the electron in the third dot interacts with the fourth dot with an energy of \(U_{g}+2V_{g}\). This results in a higher total energy than the former configuration under the parameter setting \(\alpha=0.2\).

Last, in the case of $\alpha=0.7$ with $N=L+2$, shown in Figure~\ref{fig:phase}(d), the dominating system configuration as $\varepsilon_1$ changes from $\varepsilon_1>0$ to $\varepsilon_1<0$, the configurations will appear as the following sequences:
\begin{subequations}
	\begin{alignat}{2}
		\textbf{\uppercase\expandafter{\romannumeral1}} &: &\quad |\circ,{\bullet\bullet},\bullet,{\bullet\bullet},\circ,{\bullet\bullet},\circ,...,{\bullet\bullet},\bullet,{\bullet\bullet}\rangle, \\
		\textbf{\uppercase\expandafter{\romannumeral2}} &: &\quad |\bullet, {\bullet\bullet},\bullet,{\bullet\bullet},\circ,{\bullet\bullet},\circ,...,{\bullet\bullet},\circ,{\bullet\bullet}\rangle, \\
		\textbf{\uppercase\expandafter{\romannumeral3}} &: &\quad |{\bullet\bullet}, \bullet,\bullet,{\bullet\bullet},\circ,{\bullet\bullet},\circ,...,{\bullet\bullet},\circ,{\bullet\bullet}\rangle, \\
		\textbf{\uppercase\expandafter{\romannumeral4}} &: &\quad \left|\frac{\bullet}{{\bullet\bullet}}, \circ,\bullet,{\bullet\bullet},\circ,{\bullet\bullet},\circ,...,{\bullet\bullet},\circ,{\bullet\bullet}\right\rangle, \\
		\textbf{\uppercase\expandafter{\romannumeral5}} &: &\quad \left|\frac{{\bullet\bullet}}{{\bullet\bullet}}, \circ,{\bullet\bullet},\circ,{\bullet\bullet},\circ,{\bullet\bullet},\circ,...,{\bullet\bullet},\circ\right\rangle.
	\end{alignat}
\end{subequations}

The energies corresponding to these configurations, as calculated from the Hubbard model, are as follows:  
\textbf{\uppercase\expandafter{\romannumeral1}}: $NU_{g}/2+8V_{g}$, \textbf{\uppercase\expandafter{\romannumeral2}}: $\varepsilon_{1}+NU_{g}/2+6V_{g}$, \textbf{\uppercase\expandafter{\romannumeral3}}: $NU_{g}/2+U_{g}+8V_{g}+2\varepsilon_{1}+NU_{g}/2+V_{g}+4V_{g}+2\varepsilon_{1}$, \textbf{\uppercase\expandafter{\romannumeral4}}: $3\varepsilon_{1}+NU_{g}/2+2V_{g}+2V'_{g,e}$, \textbf{\uppercase\expandafter{\romannumeral5}}: $4\varepsilon_{1}+NU_{g}/2+U_{e}+4V'_{g,e}$. Consequently, the boundaries distinguishing these regions in Figure~\ref{fig:phase}(d) can be calculated as follows:
\begin{subequations}
	\begin{alignat}{2}
		\textbf{\uppercase\expandafter{\romannumeral1}-\uppercase\expandafter{\romannumeral2}}: &\quad & \varepsilon_{1} &= 2V_{g}, \\
		\textbf{\uppercase\expandafter{\romannumeral2}-\uppercase\expandafter{\romannumeral3}}: &\quad & \varepsilon_{1} &= V_{g}, \\
		\textbf{\uppercase\expandafter{\romannumeral3}-\uppercase\expandafter{\romannumeral4}}: &\quad & \varepsilon_{1} &= 3V_{g}-2V'_{g,e}, \\
		\textbf{\uppercase\expandafter{\romannumeral4}-\uppercase\expandafter{\romannumeral5}}: &\quad & \varepsilon_{1} &= 2V_{g}-U_{e}-2V'_{g,e},
	\end{alignat}
\end{subequations}
In regions \textbf{\uppercase\expandafter{\romannumeral1}} and \textbf{\uppercase\expandafter{\romannumeral2}}, the additional electrons—two in the former and one in the latter—have the flexibility to occupy any available sites along the spin chain. In region \textbf{\uppercase\expandafter{\romannumeral3}}, a distinctive arrangement emerges where two electrons specifically occupy sites 2 and 3. This localized occupation maintains a charge density wave structure throughout the remainder of the spin chain. The region \textbf{\uppercase\expandafter{\romannumeral4}} exhibits a situation in which a single electron favors site 3, which is advantageous as it minimizes the Coulomb interaction, involving only a $2V_{\nu}$ contribution from the adjacent site 4, thereby optimizing the energy configuration. Finally, the region \textbf{\uppercase\expandafter{\romannumeral5}} naturally evolves into a global charge density wave structure, where the electron distribution systematically alternates along the entire chain, reflecting a stable and energetically favorable arrangement. This structure highlights the intrinsic properties of the system under these specific conditions.

\section{Conclusions}
\label{sec:conclusion}

In this study, we systematically explored the entanglement properties of semiconductor quantum dots within a multi-site lattice, described by the EHM. Our investigations demonstrate that local and pairwise entanglement measures respond sensitively to interactions between Coulomb forces and tunneling effects, which are influenced by the system's electronic configurations and variations in external potential energies. Notably, the entanglement characteristics show distinct phase transitions influenced heavily by coupling strength ratios and variations in potential energy. We observed that varying the potential energy of a specific dot decisively alters ground state configurations and, consequently, entanglement measures, a phenomenon that is pronounced in both weak and strong coupling regimes. This indicates that potential energy modifications can effectively control entanglement in quantum dot systems.

\section*{Acknowledgement}

We thank Ke Huang, Jiahao Wu, Quan Fu and Guo Xuan Chan
for valuable discussions. This work is supported by the Key-Area Research and Development Program of GuangDong Province  (Grant No. 2018B030326001)  the Research Grants Council of Hong Kong (Grant No. CityU 11304920),  and the Quantum Science Center of Guangdong-Hong Kong-Macao Greater Bay Area.

\end{document}